\documentclass[11pt]{article}
\usepackage[top=1in, bottom=1in, left=1in, right=1in]{geometry}
\usepackage{xcolor}
\usepackage{setspace}
\usepackage{multirow}
\usepackage{rotating}
\usepackage{amsmath,amssymb,amsthm}
\renewcommand{\qedsymbol}{$\blacksquare$}
\usepackage{comment}
\usepackage[T1]{fontenc}
\usepackage{multirow}
\usepackage{indentfirst}
\usepackage{graphicx}
\usepackage{float}
\usepackage{hhline}
\usepackage[hidelinks]{hyperref}
\usepackage{color}
\usepackage{blkarray, bigstrut}
\usepackage{array}
\newcolumntype{L}[1]{>{\raggedright\let\newline\\\arraybackslash\hspace{0pt}}m{#1}}
\newcolumntype{C}[1]{>{\centering\let\newline\\\arraybackslash\hspace{0pt}}m{#1}}
\newcolumntype{R}[1]{>{\raggedleft\let\newline\\\arraybackslash\hspace{0pt}}m{#1}}
\usepackage{amsthm}
\usepackage{amsmath} 
\usepackage{mathtools}

\usepackage{mleftright}
\usepackage[utf8]{inputenc}
\usepackage{fancyhdr}
\usepackage{breqn}
\usepackage{graphics}

\usepackage{amsmath}
\usepackage{amssymb}
\usepackage{pdflscape}
\usepackage{epstopdf}
\usepackage{lscape}
\usepackage{epsfig}
\usepackage{caption} 
\makeatletter
\renewcommand*\env@matrix[1][*\c@MaxMatrixCols c]{%
  \hskip -\arraycolsep
  \let\@ifnextchar\new@ifnextchar
  \array{#1}}
\makeatother

\renewcommand{\qedsymbol}

\title{\LARGE \bf
Conditional Waiting Time Analysis in Tandem Polling Queues
}
\author{
Ravi Suman, Ananth Krishnamurthy\\Industrial and Systems Engineering Department\\University of Wisconsin-Madison, Madison, WI-53726, USA}
\begin{document}
\begin{center}
\vspace{1in} \Large{\textbf{Conditional Waiting Time Analysis in Tandem Polling Queues}}\\
\vspace{0.1in}
\large{Ravi Suman, Ananth Krishnamurthy}\\
Department of Industrial and Systems Engineering\\
University of Wisconsin-Madison\\
1513 University Avenue, Madison, WI-53706, USA.\\
Email: rsuman@wisc.edu, ananth.krishnamurthy@wisc.edu\\
\end{center}
\begin{abstract} 
We analyze a tandem network of polling queues with two product types and two stations. We assume that external arrivals to the network follow a Poisson process, and service times at each station are exponentially distributed. For this system, we determine the mean conditional waiting time for an arriving customer using a sample path analysis approach. The approach classifies system state upon arrival into scenarios and exploits an inherent structure in the sequence of events that occur till the customer departs to obtain conditional waiting time estimates. We conduct numerical studies to show both the accuracy of our conditional waiting time estimates and their practical importance.
\end{abstract}

\section{Introduction}\label{Introduction}
A polling queue consists of a single server station serving products from different queues in a fixed order. Polling queues find applications when multiple product types compete for a common resource $\left(\text{server}\right)$. In manufacturing, polling queues have been used to model flow of multiple products undergoing manufacturing operations in a factory. In healthcare, polling queues have been used to model the flow of different types of patients through various activities in a hospital or clinic. In transportation, polling queues have been used to model multiple traffic flows in a transportation network. The design of a polling queue involves several decisions related to service discipline, queueing discipline, and buffer capacity. Takagi \cite{Takagi2000} and Vishnevskii \& Semenova \cite{Vishnevskii2006} provide a comprehensive survey of literature on the analysis of polling queues.\\

This paper focuses on tandem polling queues with multiple stations and multiple products. Our work is motivated by collaborations with manufacturing industries that were interested in obtaining better estimates of waiting times to guarantee better service to customers. We consider a tandem network of polling queues with two stations and two products, where external arrivals occur at station 1 and customers visit station 2 after completing service at station 1. For this system, we are interested in estimating the mean waiting time for an arriving customer, conditioned on the state of the network seen by the arriving customer, denoted by $W^{a}$. These conditioned mean waiting times are hard to compute in polling systems due to interaction between products and stations. However, in many applications, these conditional mean waiting times can be very different from the mean waiting times seen by an arbitrary customer denoted by $W$. In such cases, quantifying this difference is important from the perspective of waiting time estimates for scheduling or promising service levels to the customers. Despite the importance, there are very little studies on the estimation of conditional waiting times in polling queues. We believe our work takes a first step in addressing this gap in the literature.\\

We determine conditional waiting times using a sample path approach under Markovian settings. Our analysis recognizes that an arriving customer sees a system state that can be classified under one of four scenarios. Under each scenario, the sample path of events that occur before the arriving customer departs the network can be classified into sub-scenarios with distinct events. The conditional waiting times are then determined as a function of the probability of these events and their mean durations. While the approach seems tedious at first glance, we show that the sample path contain several sets of repeated events, enabling the computation using a simple algorithm. Further, the probabilities of various events, leverage only a few probabilistic results related to tandem queues.\\

Our contributions in this paper are as follows. We provide a systematic approach for estimation of conditional waiting times in a tandem network of polling queues; thus addressing an important gap in the existing literature. We show through numerical computations that our approach provides estimates with reasonable accuracy when compared to estimates obtained from simulation. We also show that the conditional waiting times for polling systems can be significantly different from mean waiting times that are often the focus of analysis of polling systems. We believe our findings will encourage additional research on conditional waiting times for polling system.\\

The remainder of the paper is structured as follows. Section \ref{Literature Review} provides a brief literature review on polling queues.  In Section \ref{Model Description and Approach}, we describe the model and in Section \ref{Preliminaries}, we derive preliminary results for analyzing the polling queues. Next, we explain the derivation of mean conditional waiting times in Section \ref{WaitingTimes}. The numerical results have been presented in Section \ref{NumericalAnalysis}. In Section \ref{Extensions}, we describe possible extensions to the method. Finally, we conclude in Section \ref{Conclusions} with final remarks.

\section{Literature Review}\label{Literature Review}
Polling queues and their applications have been an active field of research for the past few decades. Takagi et al. \cite{Takagi2000}, Vishnevskii et al. \cite{Vishnevskii2006}, and  Boona et al. \cite{Boona11} provide a comprehensive survey on polling queues and their applications. In this section, we group our discussion of the literature on polling queues in three categories$\colon$vacation models, transform based models, and mean-value analysis.\\

\textbf{Vacation Models}$\colon$ One of the earliest techniques for analyzing polling queues uses a \emph{server vacation model}, where the server periodically leaves a queue and takes a vacation to serve other queues. Fuhrmann et al. \cite{Fuhrmann85} uses such a vacation model to study a symmetric polling station with $Q$ queues served in a cyclic order by a single server and determines the expressions for sojourn times under exhaustive, gated, and $k$-limited service discipline. They show that the stationary number of customers in a single station polling queue (summed over all the queues) can be written as the sum of three independent random variables$\colon\left(i\right)$ the stationary number of customers in a standard M/G/I queue with a dedicated server, $\left(ii\right)$ the number of customers in the system when the server begins an arbitrary vacation (changeover), and $\left(iii\right)$ number of arrivals in the system during the changeover. Bertsimas et al. \cite{Bertsimas99} extend this analysis to polling systems under dynamic polling order.\\

Boxma et al. \cite{Boxma87} use a stochastic decomposition to estimate the amount of work (time needed to serve a specific number of customers) in cyclic-service systems with hybrid service strategies (e.g., semi-exhaustive for first product class, exhaustive for second and third product class, and gated for remaining product classes) and use the decomposition results to obtain a pseudo-conservation law for such cyclic systems. Cooper et al. \cite{RBCooper96} propose a decomposition theorem for polling queues with non-zero switchover times and show that the mean waiting times is the sum of two terms$\colon\left(\text{1}\right)$ the mean waiting time in a "corresponding" model in which the switchover times are zero, and $\left(\text{2}\right)$ a simple term that is a function of mean switchover times. Resing \cite{Resing93} determines the distribution for the number of customers in a queue in a single-station polling queue with zero or non-zero switchover times operating under exhaustive and gated service disciplines.\\

\textbf{Transform Based Models}$\colon$ Several studies have used transform methods to find the distributions for waiting times, cycle times, and queue lengths in a single-station polling queue. Srinivasan et al. \cite{Srinivasan95} derives a relation between Laplace–Stieltjes Transform $\left(\text{LST}\right)$ of the waiting times in a polling queue with zero-switchover-times model and a polling queue with nonzero-switchover-times for exhaustive and gated service. This relation allows one to compute the moments of the waiting times in a polling queue with nonzero-switchover-times using a simple algorithm. Borst et al. \cite{Borst97} derives the joint queue length distribution at beginning and end times of a visit to a queue in polling systems with zero and non-zero switchovers. Boxma et al. \cite{Boxma11} derive the generating function and LST of the time-stationary joint queue length and workload distributions for a single polling station with multiple-queues. Boxma et al. \cite{Boxma09} analyzes a polling system with Q-queues operating under gated policy with non-zero switchover times and determine the LST for cycle times under different scheduling disciplines. They show that LST of cycle times is only dependent on the polling discipline at each queue and is independent of the scheduling discipline used within each queue.\\

\textbf{Mean Value Analysis}$\colon$ Winands et al. \cite{Winands06} calculates the mean waiting times in a single-station multi-class polling queue for both exhaustive and gated service disciplines. They use mean value analysis to determine the mean waiting times at the polling queue for cases with non-zero setup times. Winands \cite{Winands11} presents an exact asymptotic analysis of waiting times in polling systems with exhaustive and gated disciplines when the setup times tend to infinity. This analysis helps in understanding the behavior of the polling system with large setup times.\\

Our research in this paper addresses two main gaps in the literature. First, most of the existing literature focus on analyzing single station polling queues, and there are very few studies that analyze a network of tandem polling queues. Our work is an initial step towards the analysis of systems with more than one polling stations. Second, there are no prior studies in the literature that determine the mean conditional waiting time for specific orders in polling queues. The existing literature focuses on either waiting time distributions or mean waiting time of an arbitrary customer. Our paper focuses on the mean waiting time a customer is likely to experience conditioned on the state they see upon arrival. Such conditional waiting time for a given customer can be more important for service expectations than other waiting time measures. To address these gaps, in this paper, we analyze a tandem polling queue with exhaustive service policy and determine the mean conditional waiting times using a sample path approach.
\section{Model Description and Approach}\label{Model Description and Approach}
In this section, we define the tandem polling network under consideration and provide high level description of the approach used to determine the conditional waiting times.

\subsection{Model Description}\label{Model Description}
\begin{figure}[h]
\begin{center}
\includegraphics[scale=0.30]{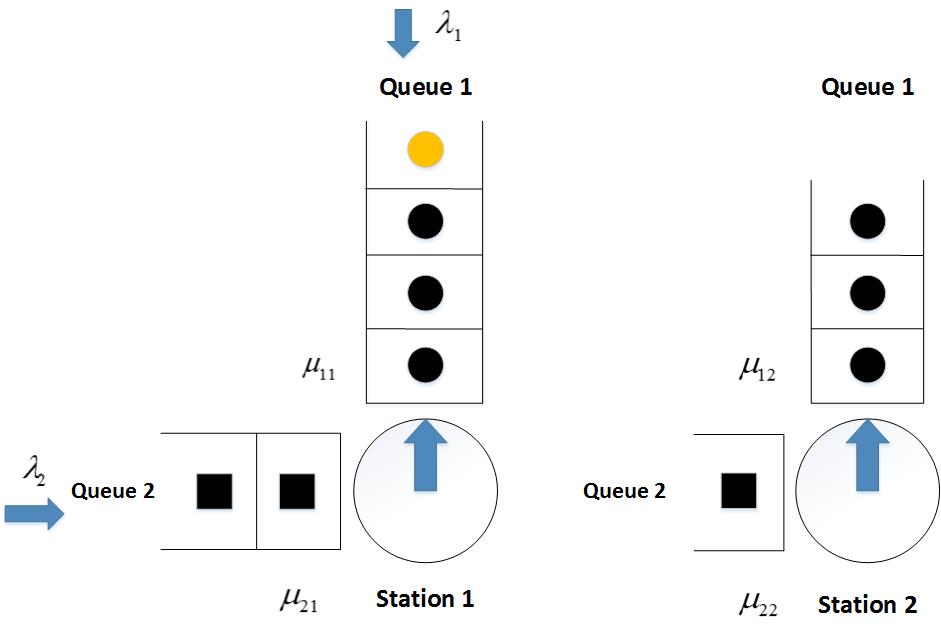}
\caption{Tandem network of polling queues with two products and two stations.}
\label{fig:mesh3.1}
\end{center}
\end{figure}

We consider a system of tandem polling queues with two stations $\left(j={}1, 2\right)$, and two types of customer class $\left(i={}1, 2\right)$ $\left(\text{see Figure } \ref{fig:mesh3.1}\right)$. Customers of type $i$ arrive to their respective queue at station 1 according to Poisson process with parameter $\lambda_i$ and have service time with exponential distribution having parameter $\mu_{i1}$ and $\mu_{i2}$ at station 1 and 2 respectively. We define $\tau_{ij}$ $\left(={}1/\mu_{ij}\right)$ as the mean service time of customer \emph{i} at station \emph{j}. Each customer first gets served at station 1 and then at station 2 before exiting the system. At each station, the server uses a cyclic discipline with exhaustive service to serve customers of type 1 and type 2 alternatively. We assume that the buffers are infinite and that switchover time is negligible at each station. The state of the system at time $t$ is defined by $\Big(\,L_{11}\left(t\right), L_{21}\left(t\right), S_{1}\left(t\right), L_{12}\left(t\right), L_{22}\left(t\right), S_{2}\left(t\right)\,\Big)$ where $L_{ij}\left(t\right)$ is the number of customers of type $i$ at station $j$ at time $t$ and $S_{j}(t)$ is the queue being served at station $j$ at time $t$. Also, let $\boldsymbol{L^{a}}={}\left(\,L_{11}\left(0\right), L_{21}\left(0\right), L_{12}\left(0\right), L_{22}\left(0\right)\,\right)$ denote the queue length seen by an arriving customer at $t={}0$.\\

\graphicspath {{Figures/}}

The traffic intensity $\rho_{ij}$ at queue \emph{i} of station \emph{j} is defined as $\rho_{ij}={}\lambda_i\tau_{ij}$ and the total traffic intensity at  station \emph{j}, $\rho_{j}$, is defined as $\rho_{j}={}\sum_{i={}1}^{2} \rho_{ij}$. Then, from Takagi et al. \cite{Takagi90}, for the system to be stable, it is required that  $\rho_{j} < $ 1 for each $j$. We assume this condition holds for our system. Let $W_{ij}^{a}$ denote the conditional waiting times for arriving type $i$ customer at station $j$ seeing a system with queue lengths $\boldsymbol{L^{a}}$ and let  $W_{i}^{a}$ denote the conditional waiting times for arriving type $i$ customer in the system seeing a system with queue lengths $\boldsymbol{L^{a}}$. Then, $\displaystyle\mathop{\mathbb{E}}\left[W_{i}^{a}\right]={}\displaystyle\mathop{\mathbb{E}}\left[W_{i1}^{a}\right] + \displaystyle\mathop{\mathbb{E}}\left[W_{i2}^{a}\right]$ and our goal in this paper is to determine $\displaystyle\mathop{\mathbb{E}}\left[W_{i}^{a}\right]$ for $i={}1, 2$.\\

\begin{figure}[h]
\begin{center}
\includegraphics[scale=0.18]{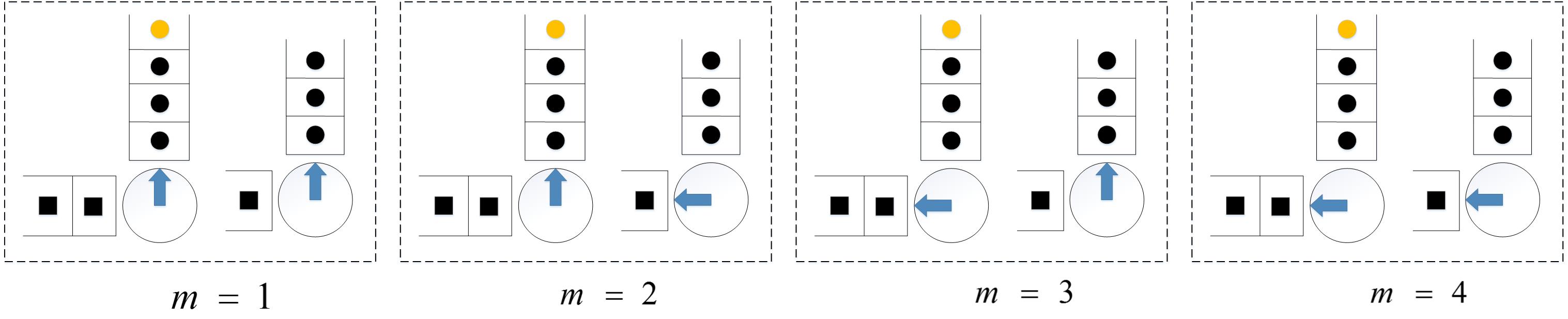}
\caption{Scenarios on the arrival of the tagged customer.}
\label{fig:mesh3.2}
\end{center}
\end{figure}

Without the loss of generality, we describe the estimation of waiting times assuming that the tagged customer corresponds to type 1. Let $L_{ij}^{'}\left(t\right)$ be the number of customers ahead of the tagged customer at station $j$ at time $t$. When the tagged customer arrives at queue 1  of station 1, the servers at station 1 and 2 could either be serving queue 1 or 2. Correspondingly, four scenarios $\left(\emph{m} ={}\text{1, \ldots, 4}\right)$ are possible depending on whether $S_{j}\left(0\right)={}$ 1 or 2 as shown in Figure $\ref{fig:mesh3.2}$. Let $m=1$ correspond to the case where $S_{1}\left(0\right)={}S_{2}\left(0\right)={}1$ and let $m={}2$ correspond to the case where $S_{1}\left(0\right)={}1$, $S_{2}\left(0\right)={}2$. Similarly, let $m={}3$ correspond to the case where $S_{1}\left(0\right)={}2$, $S_{2}\left(0\right)={}1$, and $m={}4$ correspond to the case where $S_{1}\left(0\right)={}S_{2}\left(0\right)={}2$ respectively. For each scenario $m$, $\displaystyle \mathop{\mathbb{E}}\left[W_{i}^{a}| m\right]$ is defined as the mean conditional waiting times for an arriving customer, conditioned on scenario $m$.\\
\graphicspath {{Figures/}}

To determine $\displaystyle\mathop{\mathbb{E}}\left[W_{i}^{a} | m\right]$ for an arriving customer in the system, we consider different sub-scenarios $\left(n = {}\text{1, 2, \ldots}\right)$ for each of the scenarios $m$. A sub-scenario is defined as sequence of events required for an arriving tagged customer to exit the system from station 2. Then, to determine $\displaystyle\mathop{\mathbb{E}}\left[W_{i}^{a} | m\right]$, we first determine the mean conditional waiting times $\displaystyle\mathop{\mathbb{E}}\Big[W_{i}^{a}| m, n\Big]$ for each sub-scenarios under scenario $m$, and then determine the conditional expected waiting times $\displaystyle \mathop{\mathbb{E}}\left[W_{i}^{a}| m\right]$ using Equation $\left(\ref{eq:3.1}\right)$ below$\colon$
\begin{equation}\label{eq:3.1}
\displaystyle\mathop{\mathbb{E}}\left[W_{i}^{a}| m\right]={}\sum_{s={}1}^{\infty}\displaystyle \mathop{\mathbb{E}}\left[W_{i}^{a}| m, n={}s\right]\Pr\left(n={}s| m\right)
\end{equation}
where $\Pr\left(n={}s| m\right)$ is the probability of sub-scenarios under scenario $m$. The details of the analysis of these sub-scenarios are provided in Section \ref{WaitingTimes}. However, it is useful to make an important observation regarding this analysis here. The events that constitute the various sub-scenarios for each scenarios are summarized in Table \ref{T:3.1}. We simplify our analysis by leveraging the occurrence of repeating events across sub-scenarios. For instance, in sub-scenarios $n \geq 5$ for scenario $m={}1$, the sequence of events begin to repeat. Therefore, we use the analysis of events in sub-scenario $n={}4$ to analyze the events in the sub-scenarios $n \geq 5$. Additionally, for different sub-scenarios under scenarios $m={}2, 3,$ and 4, either portions of sequence of event are similar to the sequence of events observed for sub-scenarios under scenario $m={}1$ or have some additional events than the sequence of events observed for sub-scenarios under scenario $m={}1$ $\left(\text{see highlights in Table $\ref{T:3.1}$}\right)$. We use this information to simplify our analysis of scenarios  $m={}2, 3,$ and 4.\\

\renewcommand{\arraystretch}{1.25}
\begin{table}[h!]
\centering
\caption{List of sub-scenarios for different scenarios.}\label{T:3.1}
\scalebox{0.8}{\begin{tabular}{ |c|l||c|l| }
\hline
\multicolumn{2}{|c||}{$m={}1$} & \multicolumn{2}{c|}{$m={}2$}\\
\hline
Sub-scenarios & Event & Sub-scenarios & Event\\ 
\hline
 $n = {}1   $ & A  $\prec$ B &  &\\
 $n = {}2   $ & A' $\prec$ C $\prec$   D $\prec \text{E}_{1}$ &  $n = {}1$ & C $\prec$   D $\prec \text{E}_{1}$\\
 $n = {}3   $ & A' $\prec$ C' $\prec   \text{F}_{1}  \prec \text{E}_{2}$ & $n = {}2  $ &  C' $\prec \text{F}_{1}  \prec \text{E}_{2}$\\
 $n = {}4   $ & A' $\prec$ C' $\prec   \text{F'}_{1} \prec$ G $\prec$ H &  $n = {}3   $ & C' $\prec   \text{F'}_{1} \prec$ G $\prec$ H\\
  ...            &  ...  &  ...  &  ... \\ 
\hline
\hline
\multicolumn{2}{|c||}{$m={}3$} & \multicolumn{2}{c|}{$m={}4$}\\
\hline
Sub-scenarios & Event & Sub-scenarios & Event\\ 
\hline
$n = {}1   $ &  \color{green}{J}  \color{black} $\prec$ A  $\prec$ B & $n = {}1   $ &  \color{blue} L \color{black} $\prec$ J  $\prec$ A  $\prec$ B\\ 
$n = {}2   $ &  \color{green}{J}  \color{black} $\prec$ A' $\prec$ C $\prec$   D $\prec \text{E}_{1}$ & $n = {}2   $ & \color{blue} L  \color{black} $\prec$  J $\prec$ A' $\prec$ C $\prec$   D $\prec \text{E}_{1}$\\ 
$n = {}3   $ &  \color{green}{J}  \color{black} $\prec$ A' $\prec$ C' $\prec   \text{F}_{1}  \prec \text{E}_{2}$ & $n = {}3   $ &  \color{blue} L \color{black} $\prec$  J $\prec$ A' $\prec$ C' $\prec   \text{F}_{1}  \prec \text{E}_{2}$\\ 
$n = {}4   $ &  \color{green}{J}  \color{black} $\prec$ A' $\prec$ C' $\prec   \text{F'}_{1} \prec$ G $\prec$ H & $n = {}4   $ & \color{blue} L \color{black} $\prec$ J $\prec$ A' $\prec$ C' $\prec   \text{F'}_{1} \prec$ G $\prec$ H\\  
 ...  &  \hspace{1.0cm}  ... & $\vdots$ &  \hspace{1.2cm} $\vdots$ \\
$l = {}1    $ &  \color{brown}{J'} $\prec$ K  \color{black} $\prec$ C  $\prec$   D $\prec \text{E}_{1}$ & $l = {}1   $ &  \color{red} L' \color{black} $\prec$ C  $\prec$   D $\prec \text{E}_{1}$\\ 
$l = {}2    $ &  \color{brown}{J'} $\prec$ K  \color{black} $\prec$ C' $\prec   \text{F}_{1}  \prec \text{E}_{2}$ & $l = {}2    $ &  \color{red} L' \color{black} $\prec$ C' $\prec   \text{F}_{1}  \prec \text{E}_{2}$\\
$l = {}3    $ &  \color{brown}{J'} $\prec$ K  \color{black} $\prec$ C' $\prec   \text{F'}_{1} \prec$ G $\prec$ H & $l = {}3    $ & \color{red} L' \color{black} $\prec$ C' $\prec   \text{F'}_{1} \prec$ G $\prec$ H\\ 
...  &  \hspace{1.0cm} ... &  $\vdots$ & \hspace{1.2cm} $\vdots$\\
\hline
\end{tabular}}
\end{table}
\normalsize

\section{Preliminaries}\label{Preliminaries}
Before we determine $\displaystyle\mathop{\mathbb{E}}\left[W_{i}^{a}| m\right]$ for each scenarios in Section \ref{WaitingTimes}, in this section, we present seven results that will be used multiple times in the analysis. In Result \ref{Result1}, we state the expression for hitting time to 0 for an M/M/1 queue. In Result \ref{Result2}, we consider a tandem queue setting with external arrivals only at station 1 and determine the probability that station 2 empties before station 1. For this same setting, in Result \ref{Result3}, we determine the hitting time to 0 for station 2. Next, in Result \ref{Result4}, we consider a tandem polling queue and derive the expression for the probability that exactly $k$ customers are served at station 1 given that $w$ of the same type were completely served at station 2. Result \ref{Result5} analyzes a tandem queue network and lists the expressions for waiting time for a tagged customer queued at station 1 before it exits the system after getting served at station 2. In Result \ref{Result6}, we consider two independent queues and determine the probability of $u$ service completions at one station before $w$ service completions at the other station. Finally, in Result \ref{Result7}, for a system of two independent queues, but external arrivals only at station 1, we estimate the probability that station 2 empties before station 1. The details are in the paragraphs below.\\

\theoremstyle{definition}\newtheorem{rant}{Angry Rant}
\newtheorem{result}{Result}
Consider an M/M/1 queue as shown in Figure \ref{fig:mesh3.3} below. The hitting time to 0 since time \emph{t} is defined as the measure of the time until the instant, a customer departs leaving behind an empty system for the first time.
\graphicspath {{Figures/}}
\begin{figure}[h]
\begin{center}
\includegraphics[scale=0.40]{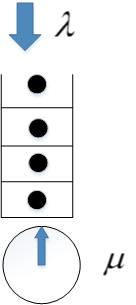}
\caption{Single station M/M/1 queue.}
\label{fig:mesh3.3}
\end{center}
\end{figure}
\begin{result}\label{Result1}
Let $g\left(L\left(t\right)\right)$ be the hitting time to 0 since \emph{t} for an M/M/1 queue $\left(\text{see Figure \ref{fig:mesh3.3}}\right)$ with Poisson arrivals with parameter $\lambda$, exponential service times with parameter $\mu$, and $L\left(t\right)$ customers at \emph{t}. Then, the mean value of  $g\left(L\left(t\right)\right)$ is $\frac{L\left(t\right)}{\left(\mu-\lambda\right)}$ and the explicit expression for the probability distribution function $f_{g\left(L\left(t\right)\right)}$ of the random variable $g\left(L\left(t\right)\right)$ has been derived in Prabhu et al. \cite{Prabhu60} and reproduced below as Equation $\left(\ref{eq:3.2}\right)$.
\begin{equation}\label{eq:3.2}
f_{g\left(L\left(t\right)\right)}\left(t\right) ={} \frac{L\left(t\right)}{t}e^{-\left(\lambda+\mu\right)t}\sqrt[\leftroot{-2}\uproot{2} L\left(t\right)]{\frac{\mu}{\lambda}}I_{L\left(t\right)}\left(2t\sqrt{\lambda\mu}\right)\hspace{2cm}\text{for } t > 0
\end{equation}
where $I_{L\left(t\right)}\left(x\right)$ is the modified Bessel function defined as$\colon$

\begin{equation}\label{eq:3.3}
I_{L\left(t\right)}\left(x\right)={}\sum_{k={}0}^{\infty}\frac{\left(x/2\right)^{2k + L\left(t\right)}}{k!\left(k + L\left(t\right)\right)!}
\end{equation}$\hfill\square$
\end{result}

\begin{figure}[h]
\begin{center}
\includegraphics[scale=0.50]{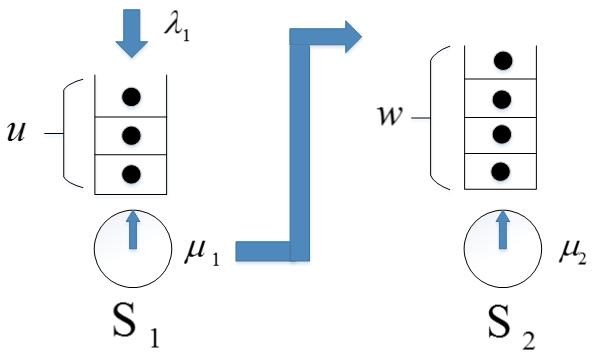}
\caption{Two station tandem queueing system.}
\label{fig:mesh3.4}
\end{center}
\end{figure}

In the next result, we analyze hitting times in a tandem queueing system. Consider a two station tandem queueing system $\left(\text{see Figure \ref{fig:mesh3.4}}\right)$ with Poisson arrivals at station 1 and exponential service times at the two stations.\\

\graphicspath {{Figures/}}

Assume that customers arriving at station 1 get served at station 2 and then exit the system, and that stations follow a FCFS policy. The state of the system at any instant is defined by $\left(i, j\right)$, where $i$ and $j$ denote the number of customers at station 1 and 2 respectively and let set $\mathcal{S}$ contain all the states of the underlying continuous Markov process $X\left(t\right)$ describing the state transitions of this system and let the state of the system at $t={}0$ is $\left(u, w\right)$. We state two results, namely Result \ref{Result2} and  Result \ref{Result3} for this system.
\begin{result}\label{Result2}
Let $p_{1}\left(u, w\right)$ be the probability that station 1 empties before station 2 and let $p_{2}\left(u, w\right)$ be the probability that station 2 empties before station 1. Then, for $k > 0\colon$
\begin{align*}
p_{1}\left(u, w\right) & = {}1 - \sum_{k={}1}^{\infty}\alpha_{\left(k, 0\right)}\left(u, w\right)\\
p_{2}\left(u, w\right) & = {}\sum_{k={}1}^{\infty}\alpha_{\left(k, 0\right)}\left(u, w\right)
\end{align*}
where $\alpha_{\left(k, 0\right)}$ is a function defined using the entries in the underlying Markov chain with transition matrix $P$.
\end{result}

\begin{proof}
See appendix \ref{AppendixA}.$\hfill \square$\\
\end{proof}
Next, in Result \ref{Result3}, for the tandem queuing system described in Figure \ref{fig:mesh3.4}, we determine the expected time $\phi_{\left(u, w\right)}$ for station $x$ to become empty, given that station $x$ becomes empty before station $y$, where $x$ and $y$ take values 1, 2, and $x \neq {}y$.

\begin{result}\label{Result3}
The expected time $\phi_{\left(u, w\right)}$ is approximated by the solution to the following system of equations$\colon$
\begin{align}
    \begin{cases}
      \phi_{\left(u, w\right)} = {} 0 \hspace{6.93cm} \left(u, w\right) \in \mathcal{R}_{x}\nonumber\\
      \phi_{\left(u, w\right)} = {} 1 + \sum_{\left(i, j\right)} Q_{\left(u, w\right)\left(i, j\right)}\phi_{\left(i, j\right)} = {}0  \hspace{2.2cm} \left(u, w\right) \notin \mathcal{R}_{x}\\
    \end{cases}
\end{align}
where $Q_{\left(u, w\right)}$ is a sub-matrix of the transition matrix $P$.
\end{result}

\begin{proof}
See appendix \ref{AppendixA}.$\hfill \square$\\
\end{proof}

In Result \ref{Result4}, we determine probability of service completions at stations in a tandem polling queue.\\

\graphicspath {{Figures/}}
\begin{figure}[H]
\begin{center}
\includegraphics[scale=0.50]{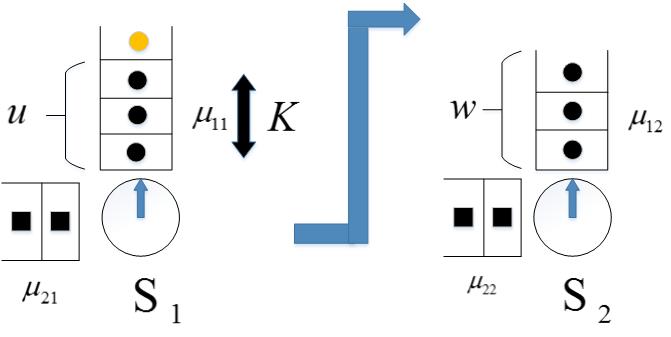}
\caption{Tandem polling queue with two stations.}
\label{fig:mesh3.5}
\end{center}
\end{figure}

Consider a two single server stations $\left(\text{see Figure \ref{fig:mesh3.5}}\right)$ with $u$ and $w$ type 1 customers at stations 1 and 2 respectively at $t={}0$. Let the service times be exponential with mean $\mu_{i1}^{-1}$ and $\mu_{i2}^{-1}$ at stations 1 and 2 respectively for type $i$ customers. There are no external arrival of customers and customers served at station 1 move to station 2 for service. Assume the server at station 2 switches from serving queue $i$ during its first cycle after serving $w + K$ customers since the arrival of the tagged customer at station 1, where $K$ is a non-negative random variable taking values $0, 1, 2$ and represents the number of customers that transferred from station 1 to station 2 and were served in the first cycle at station 2.
\begin{result}\label{Result4}
 The probability $\displaystyle \mathop{\mathbb{P}}\Big(K_{i}={}k\Big)$, that exactly $k$ type $i$ customers are served at station 1 when $w$ type $i$ customers are completely served at station 2 is given by$\colon$
\begin{equation}\label{eq:3.4}
\displaystyle \mathop{\mathbb{P}}\Big(K_{i}={}k\Big)= {}\left(\frac{\mu_{i1}}{\mu_{i1} + \mu_{i2}}\right)^k \left(\frac{\mu_{i2}}{\mu_{i1} + \mu_{i2}}\right)^{w+k}\left[\binom{2k + w -1}{k} - \binom{2k + w -1}{k-1}\right]
\end{equation}
\end{result}

\begin{proof}
See appendix \ref{AppendixA}.$\hfill \square$\\
\end{proof}

Next, in Result \ref{Result5}, we determine the waiting time for a tagged customer in a tandem queueing network.\\

\graphicspath {{Figures/}}
\begin{figure}[H]
\begin{center}
\includegraphics[scale=0.50]{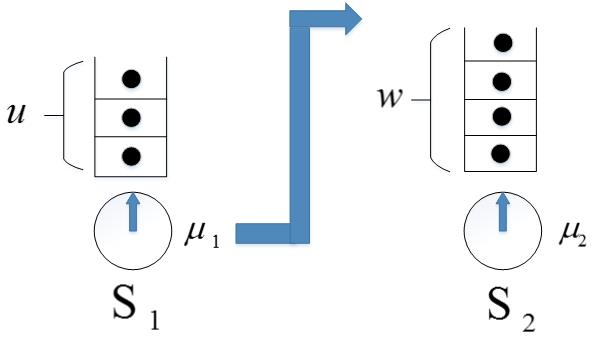}
\caption{Tandem queues with two single server stations.}
\label{fig:mesh3.6}
\end{center}
\end{figure}
Consider a two single server station $\left(\text{see Figure \ref{fig:mesh3.6}}\right)$ with $u$ customers at station 1 and $w$ customers at station 2 at $t={}0$ and exponential service times with mean $\mu_{1}^{-1}$ and $\mu_{2}^{-1}$ respectively. There are no external arrival of customers to both the stations and customers served at station 1 move to station 2 for service.

\begin{result}\label{Result5}
Let $W_{u, w}$ be the waiting time for the tagged customer queued at station 1 before it exits the system after getting served at station 2. Then$\colon$
\begin{align}
\displaystyle \mathop{\mathbb{E}}\left[W_{u, w}\right]={}
    \begin{cases}
       \mu_{1}^{-1} + \displaystyle\mathop{\mathbb{E}}\left[W_{u-1, 1}\right]\hspace{5.88cm} \text{ if } \left(u > 0, w ={}0\right)\\
       p\Big(\mu_{1}^{-1} + \left(w+1\right)\mu_{2}^{-1}\Big) + q\Big(\displaystyle\mathop{\mathbb{E}}\left[W_{0, w-1}\right]\Big)\hspace{2.0cm} \text{ if } \left(u = 0, w >0\right)\nonumber\\
       p\Big(\mu_{1}^{-1}+ \displaystyle\mathop{\mathbb{E}}\left[W_{u-1, w+1}\right]\Big) + q\Big(\displaystyle\mathop{\mathbb{E}}\left[W_{u, w-1}\right]\Big) \hspace{1.7cm} \text{ if } \left(u > 0, w > 0\right)\\
    \end{cases}
\end{align}
\end{result}
\vspace{3mm}
\begin{proof}
See appendix \ref{AppendixA}.$\hfill \square$\\
\end{proof}

In Result \ref{Result6}, we determine the probability of service completions at two stations.\\

\graphicspath {{Figures/}}
\begin{figure}[H]
\begin{center}
\includegraphics[scale=0.50]{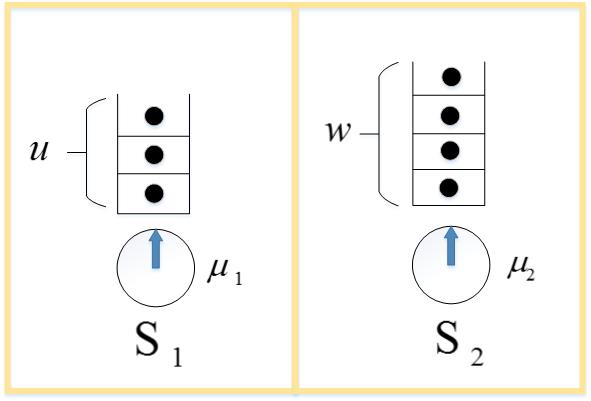}
\caption{Operations of two independent single server queues.}
\label{fig:mesh3.7}
\end{center}
\end{figure}

Consider two independent single server stations 1 and 2 $\left(\text{see Figure \ref{fig:mesh3.7}}\right)$ with exponential service times with parameter $\mu_{1}$ and $\mu_{2}$ respectively, having $u$ customers at station 1 and $w$ customers at station 2 at some time $t$. 

\begin{result}\label{Result6}
Let $h_{1}\left(u\right)$ be the time to serve the $u$ customers at station 1 and $h_{2}\left(w\right)$ be the time to serve $w$ customers at station 2. Then $h_{1}\left(u\right)$ and $h_{2}\left(w\right)$ are random variables with gamma distribution with means $\frac{u}{\mu_{1}}$ and $\frac{w}{\mu_{2}}$ respectively. Subsequently for $\left(u,\text{ }w\right) \in \displaystyle\mathop{\mathbb{Z^{+}}}\colon$
\begin{equation}\label{eq:3.5}
\displaystyle \mathop{\mathbb{P}}\Big(h_{1}\left(u\right) < h_{2}\left(w\right)\Big) ={}\sum_{r={}u}^{\infty} \left(\frac{\mu_{1}}{\mu_{1} + \mu_{2}}\right)^r \left(\frac{\mu_{2}}{\mu_{1} + \mu_{2}}\right)^{w} \binom{r + w -1}{r}
\end{equation}
\end{result}
\begin{proof}
See appendix \ref{AppendixA}.$\hfill \square$\\
\end{proof}
Finally, in Result \ref{Result7}, we estimate the probability that station 2 empties before station 1 given no external arrivals at station 1.\\
\graphicspath {{Figures/}}
\begin{figure}[H]
\begin{center}
\includegraphics[scale=0.50]{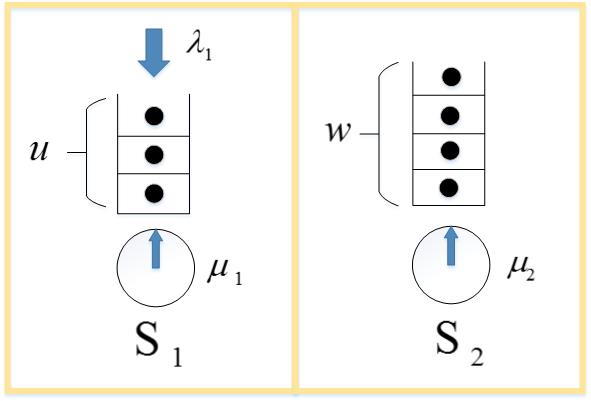}
\caption{Operations of two independent single server stations with Poisson arrivals at station 1.}
\label{fig:mesh3.8}
\end{center}
\end{figure}
Consider two independent single server stations 1 and 2 $\left(\text{see Figure \ref{fig:mesh3.8}}\right)$ having $u$ customers at station 1 and $w$ customers at station 2 at $t={}0$. Let station 1 have Poisson arrivals with parameter $\lambda_{1}$ and exponential service times with parameter $\mu_{1}$ while there are no external arrivals at station 2 and has exponential service times with parameter $\mu_{2}$. Define $g\left(u\right)$ as the hitting time to 0 since \emph{t} for station 1 and $h\left(w\right)$ as the time to serve $w$ customers at station 2.
\begin{result}\label{Result7}
Given $g\left(u\right)$, $h\left(w\right)$, and $F_{h\left(w\right)}\left(t\right) = {}\int_{0}^{t}f_{h\left(w\right)}\left(x\right)dx$, the probability that station 2 empties before station 1 is given by Equation $\left(\ref{eq:3.6}\right)$.
\begin{equation}\label{eq:3.6}
\displaystyle \mathop{\mathbb{P}}\Big(h\left(w\right) < g\left(u\right)\Big) = {}\int_{0}^{\infty}\Big[\int_{0}^{t}f_{h\left(w\right)}\left(x\right)dx\Big]f_{g\left(u\right)}\left(t\right) dt
\end{equation}
\end{result}
\begin{proof}
See appendix \ref{AppendixA}.$\hfill \square$\\
\end{proof}
In Section \ref{WaitingTimes}, we use Results $\left(\ref{Result1}\right) - \left(\ref{Result7}\right)$ to derive estimates of mean conditional waiting times.
\section{Determination of Waiting Times}\label{WaitingTimes}
In this section, we derive mean conditional waiting times. We describe the derivation for scenario $m={}1$ in detail, in the main text and report the derivations for scenario $m = {}2, 3,$ and 4 in Appendix \ref{AppendixB}, \ref{AppendixC}, and \ref{AppendixD} respectively. 
\subsection{Waiting Time Analysis of Scenario$\colon m = {}1$}\label{M1}
\begin{figure}[h]
\begin{center}
\includegraphics[scale=0.40]{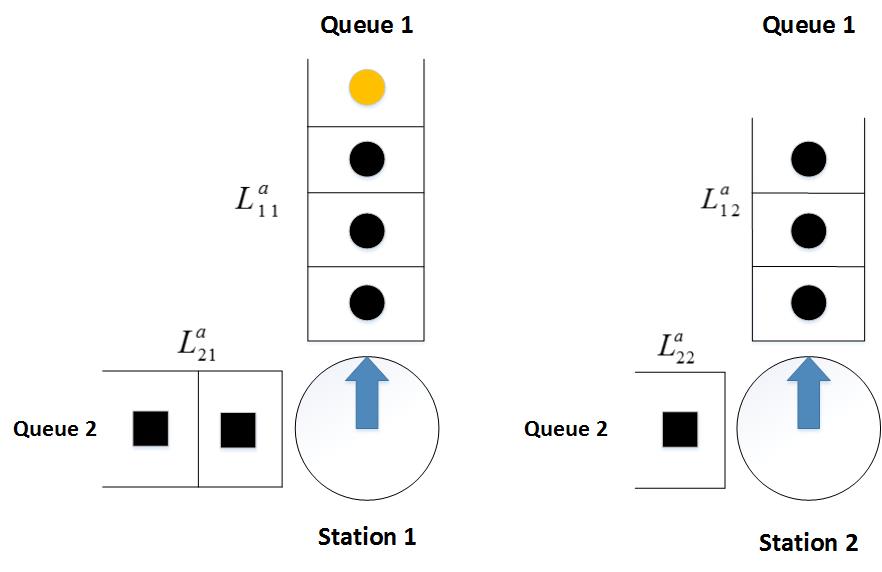}
\caption{Tandem polling queue with two stations $\left(m={}1\right)$.}
\label{fig:mesh3.9}
\end{center}
\end{figure}

In scenario $m={}1$, when the tagged customer arrives at station 1 at time $t={} 0$, the server at both the stations are serving queue 1 as shown in Figure \ref{fig:mesh3.9}. The state of the system for scenario 1 at ${t = {}0}$ can be represented as $\Big(L_{11}^{a}, L_{21}^{a}, 1, L_{12}^{a}, L_{22}^{a}, 1\Big)$ where $L_{ij}^a ={}L_{ij}\left(0\right)$. Within this scenario of $m={}1$, different sub-scenarios are possible depending on how the tagged customer gets served at station 1 and station 2. These sub-scenarios are analyzed below.\\
\graphicspath {{Figures/}}

\begin{figure}[H]
\begin{center}
\includegraphics[scale=0.20]{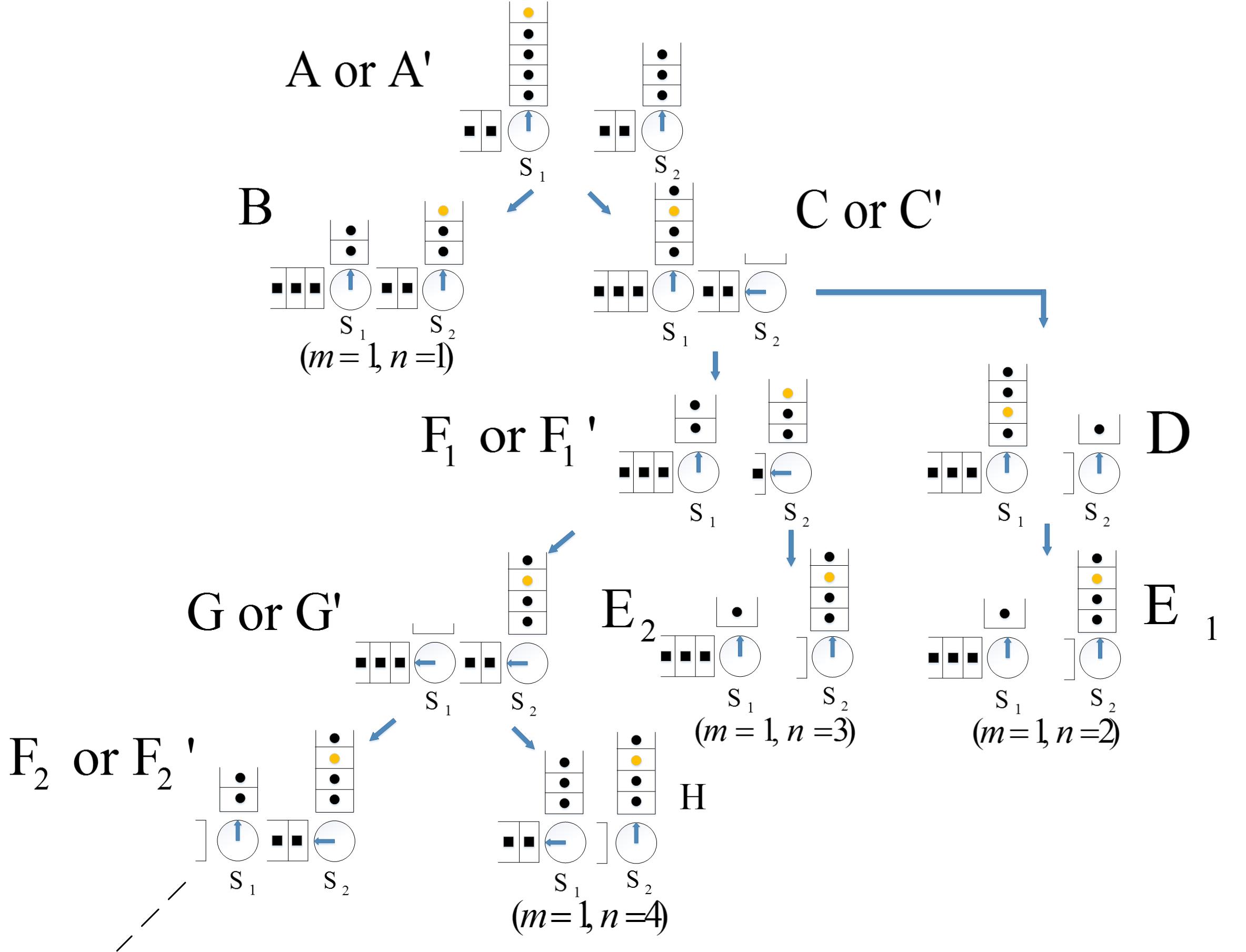}
\caption{Tree of different sub-scenarios for $m={}1$.}
\label{fig:mesh3.10}
\end{center}
\end{figure}

For $m={}1$, the different sub-scenarios are shown in Table \ref{T:3.2} along with corresponding events A, B, C, etc.\\
\renewcommand{\arraystretch}{1.00}
\begin{table}[H]
\centering
\caption{List of sub-scenarios of scenario $m={}1$.}\label{T:3.2}
\begin{tabular}{ |c|c|c| } 
\hline
Scenario & Sub-scenario & Event \\ 
\hline
$m={}1$ & $n = {}1   $ &  A  $\prec$ B\\ 
$m={}1$ & $n = {}2   $ &  A' $\prec$ C $\prec$   D $\prec \text{E}_{1}$\\ 
$m={}1$ & $n = {}3   $ &  A' $\prec$ C' $\prec   \text{F}_{1}  \prec \text{E}_{2}$\\ 
$m={}1$ & $n = {}4   $ &  A' $\prec$ C' $\prec   \text{F'}_{1} \prec$ G $\prec$ H\\ 
$m={}1$ &  ...               &  ...\\ 
\hline
\end{tabular}
\end{table}
\normalsize
The tree of sub-scenarios for scenario $m={}1$ is shown in Figure $\ref{fig:mesh3.10}$. Note that events E and F repeat in this tree but with different queue lengths and are denoted by $\text{E}_{1}$, $\text{E}_{2}$, and $\text{F}_{1}\left(\text{F'}_{1}\right)$, $\text{F}_{2}\left(\text{F'}_{2}\right)$.\\ 
\graphicspath {{Figures/}}
We first determine the conditional waiting times and probabilities for each sub-scenarios and then, calculate the waiting time $\displaystyle \mathop{\mathbb{E}}\left[W_{1}^{a}| m = {}1\right]$ using Equation $\left(\ref{eq:3.7}\right)$.
\begin{equation}\label{eq:3.7}
\displaystyle \mathop{\mathbb{E}}\left[W_{1}^{a}| m = {}1\right]={}\sum_{s={}1}^{\infty}\displaystyle \mathop{\mathbb{E}}\left[W_{1}^{a}| m = {}1, n={}s\right]\Pr\left(n={}s| m={}1\right)
\end{equation}
\subsection{Waiting Time Analysis of Sub-scenario$\colon \left(m = {}1, n = {}1\right)$}\label{M1N1}
\begin{figure}[H]
\begin{center}
\includegraphics[scale=0.30]{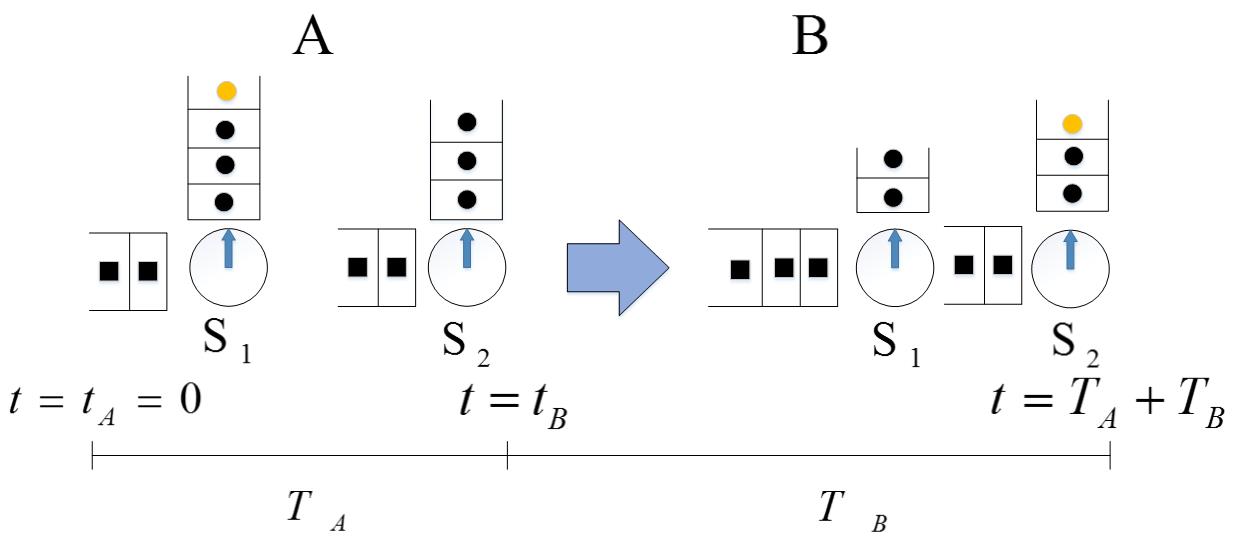}
\caption{Sequence of events for sub-scenario with $m = {}1$, $n={}1$.}
\label{fig:mesh3.11}
\end{center}
\end{figure}

From the perspective of the tagged customer arriving at station 1 at $t={}0$, the sub-scenario $\left(m = {}1, n = {}1\right)$ corresponds to a sequence of two events, A and B, that needs to occur before the tagged customer departs from station 2 $\left(\text{see Figure \ref{fig:mesh3.11}}\right)$. In the event A, for some $K_{1} > L_{11}^{a}$, exactly $K_{1}$ customers are served at station 1 before station 2 server switches to serve queue 2 at $t={}t_{B}$. As $K_{1} \geq L_{11}^{a}+1$, the tagged customer gets served at station 1 and moves to station 2 before the server at station 2 switches to queue 2.  In the event B,  the tagged customer and the $L_{11}^{a}$ customers ahead of it are served at station 2 in the first cycle of station 2. It is important to note that at no instance until the service of tagged customer at station 2 is queue 1 at station 2 empty. The probability of this sub-scenario $\displaystyle \mathop{\mathbb{P}}\Big(n ={}1| m={}1\Big)$ that $L_{11}^{a}+ 1$ type 1 customers are served at station 1 before queue 1 at station 2 becomes empty is given by Equation $\left(\ref{eq:3.8}\right)$ and is determined using Result $\ref{Result4}$.

\begin{equation}\label{eq:3.8}
 \displaystyle \mathop{\mathbb{P}}\Big(n ={}1| m={}1\Big) = {} \displaystyle \mathop{\mathbb{P}}\Big(K_{1} >  L_{11}^{a}\Big)
\end{equation}

\graphicspath {{Figures/}}

For sub-scenario $\left(m = {}1, n={}1\right)$, $\displaystyle\mathop{\mathbb{E}}\Big[W_{1}^{a}| m = {}1, n = {}1\Big]$ is the expected time needed to serve the $L_{12}^{a}$ customers at station 2 at $t={}0$ and the $L_{11}^{a} + 1$ customers that moved to station 2 after being served at station 1. For sub-scenario $\left(m = {}1, n={}1\right)$, we write $\displaystyle\mathop{\mathbb{E}}\Big[W_{1}^{a}| m = {}1, n = {}1\Big]$ as
\begin{equation}\label{eq:3.9}
\displaystyle\mathop{\mathbb{E}}\Big[W_{1}^{a}| m = {}1, n = {}1\Big]={}\displaystyle\mathop{\mathbb{E}}\Big[h_{12}\Big(L_{11}^{a} + L_{12}^{a}+ 1\Big)\Big]
\end{equation}

Next, we conduct a similar analysis for sub-scenarios $n={}2, 3, 4$ for scenario $m={}1$.

\subsection{Waiting Time Analysis of Sub-scenario$\colon \left(m = {}1, n = {}2\right)$}\label{M1N3}
\begin{figure}[h!]
\begin{center}
\includegraphics[scale=0.40]{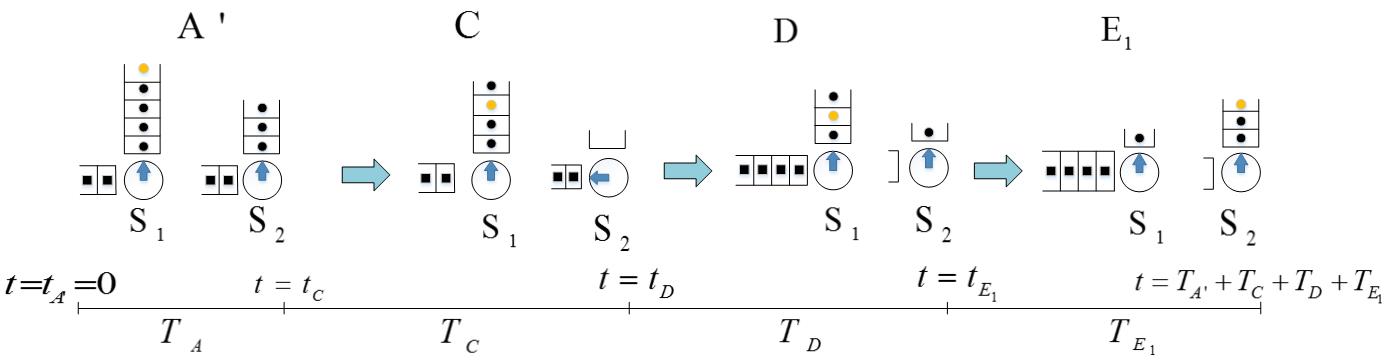}
\caption{Sequence of events for sub-scenario with $m = {}1$, $n={}2$.}
\label{fig:mesh3.12}
\end{center}
\end{figure}

From the perspective of the tagged customer arriving at station 1 at $t={} 0$, the sub-scenario $\left(m = {}1, n = {}2\right)$ corresponds to a sequence of 4 events: A', C, D, and $\text{E}_{1}$ before the tagged customer exits the system as shown in Figure $\ref{fig:mesh3.12}$.  In event A', exactly $K_{1}$ customers, where $K_{1} \leq L_{11}^{a}$, are served at station 1 before station 2 server switches to serve queue 2  at $t={}t_{C}$ after serving $ L_{12}^{a}+ K_{1}$ type 1 customers. The time $T_{A}$ it takes for server at station 2 to serve $ L_{12}^{a}+ K_{1}$ customers of queue 1 is $h_{12}\Big(L_{12}^{a}+ K_{1}\Big)$. The probability $\displaystyle \mathop{\mathbb{P}}\Big(\text{Event A'}\Big)$ is the complement of $\displaystyle \mathop{\mathbb{P}}\Big(n ={}1| m={}1\Big)$.\\

\graphicspath {{Figures/}}

In event C, all customers of queue 2 at station 2 at $t={}t_{C}$ are served before the tagged customer is served at station 1. The time $T_{C}$ it takes for the server at station 2 to serve $L_{22}\left(t_{C}\right)$ customers of queue 2 is $h_{22}\left(L_{22}\left(t_{C}\right)\right)$, where $L_{22}\left(t_{C}\right) = {} L_{22}^{a}$. The number of customers ahead of the tagged customer at $t={}t_{C}$ is
\begin{equation}\label{eq:3.10}
L_{11}^{'}\left(t_{C}\right) = {} L_{11}^{a}- K_{1}
\end{equation}
where the conditional probability of $K_{1}$ is given by Equation $\left(\ref{eq:3.12}\right)$ that uses $\displaystyle\mathop{\mathbb{P}}\Big(K_{1}={} u\Big)$ obtained using Result $\ref{Result4}$.
\begin{equation}\label{eq:3.11}
\displaystyle\mathop{\mathbb{P}}\Big(K_{1}={} k| K_{1} \leq L_{11}^{a}\Big)={}\frac{\displaystyle \mathop{\mathbb{P}}\Big(K_{1}={} v\Big)}{\sum_{k=0}^{{}L_{11}^{a}}\displaystyle \mathop{\mathbb{P}}\Big(K_{1}={} k\Big)}
\end{equation}
The probability $\displaystyle \mathop{\mathbb{P}}\Big(\text{Event C}\Big)$ that $L_{22}\left(t_{C}\right)$ customers of queue 2 at station 2 are served before $L_{11}^{'}\left(t_{C}\right)+1$ customer gets served at station 1 is given by Equation $\left(\ref{eq:3.12}\right)$  and is determined using Result $\ref{Result6}$.
\begin{equation}\label{eq:3.12}
\begin{aligned}
\displaystyle \mathop{\mathbb{P}}\Big(\text{Event C}\Big) & ={}\displaystyle \mathop{\mathbb{P}}\Big(h_{22}\Big(L_{22}\left(t_{C}\right)\Big)< {} h_{11}\left(L_{11}^{'}\left(t_{C}\right)+1\right)\Big)\\
& ={}\sum_{k={}1}^{L_{11}^{a}}\displaystyle \mathop{\mathbb{P}}\Big(h_{22}\Big(L_{22}\left(t_{C}\right)\Big)< {} h_{11}\left( L_{11}^{a}- k+1\right)\Big)\displaystyle\mathop{\mathbb{P}}\Big(K_{1}={} k| K_{1} \leq L_{11}^{a}\Big)
\end{aligned}
\end{equation}
In the events D and $\text{E}_{1}$, the tagged customer first gets served at station 1 after the service of $L_{11}^{'}\left(t_{D}\right)$ customers ahead of it and then moves to station 2 at $t={}t_{E_{1}}$, where it eventually gets served and exits the system. It is important to note that as queue 1 is being served at station 1 during event D, $L_{22}\left(t\right)$ remains 0 for $t_{D}\leq t \leq t_{E_{1}}$ and hence, the server at station 2 does not switch to serve queue 2 once it empties queue 1. Thus, $\displaystyle \mathop{\mathbb{P}}\Big(\text{Event D}\Big)\displaystyle \mathop{\mathbb{P}}\Big(\text{Event $\text{E}_{1}$}\Big)  = {}1$.\\

Let $V_{1}$ be the number of customers that got served in queue 1 at station 1 during event C. As the tagged customer did not get served at station 1 during event C, the number of customers ahead of the tagged customer at $t={}t_{D}$ is equal to $L_{11}^{'}\left(t_{D}\right) = {} {}L_{11}^{a}- K_{1} - V_{1}$ where the conditional probability distribution of $V_{1}$ is
\begin{equation}\label{eq:3.13}
\displaystyle \mathop{\mathbb{P}}\Big(V_{1}={} v| V_{1} \leq L_{11}^{a}- K_{1} \Big)={}\frac{\displaystyle \mathop{\mathbb{P}}\Big(V_{1}={} v\Big)}{\sum_{v=0}^{{}L_{11}^{a}- K_{1}}\displaystyle \mathop{\mathbb{P}}\Big(V_{1}={} v\Big)}
\end{equation}
and
\begin{equation}\label{eq:3.14}
\displaystyle \mathop{\mathbb{P}}\Big(V_{1}={} v\Big)=\Big(1/v!\Big)e^{-\mu_{11}\left(T_{C}\right)}\Big(\mu_{11}\left(T_{C}\right)\Big)^{v}
\end{equation}
Note that the number of type 1 customers at station 2 at $t={}t_{D}$, $L_{12}\left(t_{D}\right)$,  is $V_{1}$. At $t={}t_{D}$, we know that  $L_{11}^{'}\left(t_{D}\right)$ customers are ahead of the tagged customer at station 1 and $ L_{12}\left(t_{D}\right)$ customers are queued at station 2. Let $\displaystyle\mathop{\mathbb{E}}\Big[W_{L_{11}^{'}\left(t_{D}\right), L_{12}\left(t_{D}\right)}\Big]$ be the mean waiting time of the tagged customer for events D and $\text{E}_{1}$. We can determine $\displaystyle\mathop{\mathbb{E}}\Big[W_{L_{11}^{'}\left(t_{D}\right), L_{12}\left(t_{D}\right)}\Big]$ using Result \ref{Result5}.\\

Then, based on this analysis, for sub-scenario $\left(m = {}1, n={}2\right)$, $\displaystyle\mathop{\mathbb{E}}\Big[W_{1}^{a}| m = {}1, n = {}2\Big]$ is equal to the sum of the mean times for events A', C, D, and $\text{E}_{1}$ and is given by Equation $\left(\ref{eq:3.15}\right)$.
\begin{equation}\label{eq:3.15}
\begin{aligned}
\displaystyle\mathop{\mathbb{E}}\Big[W_{1}^{a}| m = {}1, n = {}2\Big] =  {}\displaystyle\mathop{\mathbb{E}}\Big[h_{12}\Big(L_{12}^{a}+ K_{1}\Big)\Big]+ \displaystyle\mathop{\mathbb{E}}\Big[h_{22}\Big(L_{22}\left(t_{c}\right)\Big)\Big] + \displaystyle\mathop{\mathbb{E}}\Big[W_{L_{11}^{'}\left(t_{D}\right), L_{12}\left(t_{D}\right)}\Big]
\end{aligned}
\end{equation}

The probability of this sub-scenario $\displaystyle \mathop{\mathbb{P}}\Big(n ={}2| m={}1\Big)$ is given by Equation $\left(\ref{eq:3.16}\right)$.
\begin{equation}\label{eq:3.16}
\displaystyle \mathop{\mathbb{P}}\Big(n ={}2| m={}1\Big) = \displaystyle \mathop{\mathbb{P}}\Big(\text{Event A'}\Big)\displaystyle \mathop{\mathbb{P}}\Big(\text{Event C}\Big)
\end{equation}
\subsection{Waiting Time Analysis of Sub-scenario$\colon \left(m = {}1, n = {}3\right)$}\label{M1N2}
\begin{figure}[h!]
\begin{center}
\includegraphics[scale=0.45]{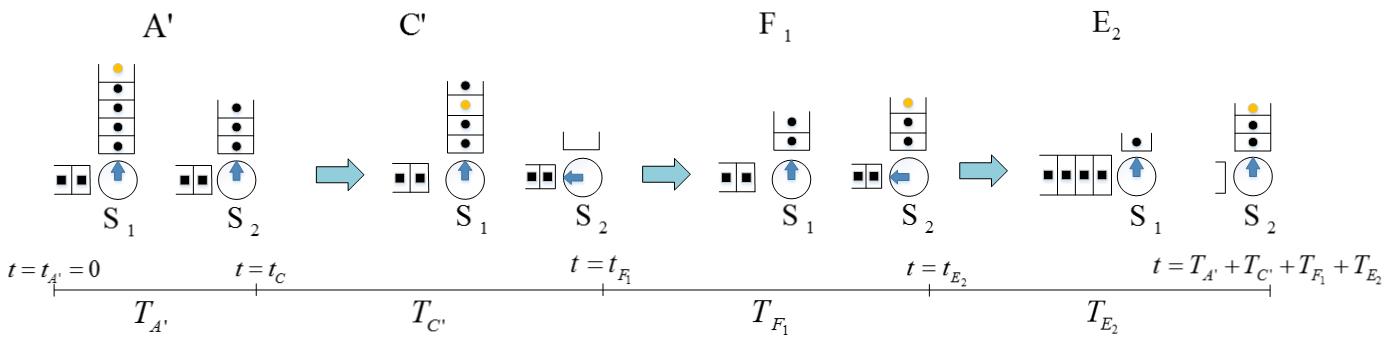}
\caption{Sequence of events for sub-scenario with $m = {}1$, $n={}3$.}
\label{fig:mesh3.13}
\end{center}
\end{figure}

From the perspective of the tagged customer arriving at station 1 at $t={}0$, the sub-scenario $\left(m = {}1, n = {}3\right)$ corresponds to a sequence of 4 events: A', C', $\text{F}_{1}$, and $\text{E}_{2}$ as shown in Figure $\ref{fig:mesh3.13}$. In event A', exactly $K_{1}$ $\Big(\text{for $K_{1} \leq L_{11}^{a}$}\Big)$ customers are served at station 1 in time $h_{12}\Big(L_{12}^{a}+ K_{1}\Big)$ before station 2 server switches to serve queue 2  at $t={}t_{C'}$ after serving $L_{12}^{a}+ K_{1}$ type 1 customers.\\

\graphicspath {{Figures/}}

In the event C', the tagged customer and the $L_{11}'\left(t_{C'}\right)$ customers ahead of it at $t={}t_{C}$ are served at station 1 before $L_{22}\left(t_{C'}\right)$ customers of queue 2 at station 2 are served. The time $T_{C'}$ it takes for server at station 1 to serve $L_{11}^{'}\left(t_{C'}\right)+1$ customers of queue 1 is $h_{11}\left(L_{11}^{'}\left(t_{C'}\right)+1\right)$, where $L_{11}^{'}\left(t_{C'}\right)$ is given by Equation $\left(\ref{eq:3.10}\right)$. The probability $\displaystyle \mathop{\mathbb{P}}\Big(\text{Event C'}\Big)$ that $L_{11}^{'}\left(t_{C'}\right)+1$ are served at station 1 before $L_{22}\left(t_{C'}\right)$ customers are served at station 2 is given by Equation $\left(\ref{eq:3.17}\right)$ and is determined using Result $\ref{Result6}$.
\begin{equation}\label{eq:3.17}
\begin{aligned}
\displaystyle \mathop{\mathbb{P}}\Big(\text{Event C'}\Big)={}\displaystyle \mathop{\mathbb{P}}\Big(h_{11}\left(L_{11}^{'}\left(t_{C'}\right)+1\right) < h_{22}\Big(L_{22}\left(t_{C'}\right)\Big)\Big)
\end{aligned}
\end{equation}

In the event $F_{1}$, all customers of queue 2 at station 2 at $t={}t_{F_{1}}$ are served before queue 1 at station 1 empties.  Let $S_{1}\left(t\right)$ be the Poisson random variable denoting number of arrivals of type 1 customers at station 1 in time $t$. During the event A' and C', external  $S_{1}\left(T_{A'} + T_{C'}\right)$ type 1 customers arrive at station 1. As the tagged customer and all the customers ahead of it at station 1 has moved to station 2 at the end of event C', the length of queue 1 at station 1 at $t={}t_{F_{1}}$ is given by Equation $\left(\ref{eq:3.18}\right)$.
\begin{equation}\label{eq:3.18}
L_{11}\left(t_{F_{1}}\right) = {}S_{1}\left(T_{A'} + T_{C'}\right) 
\end{equation}
Let $V_{2}$ be the number of customers that got served in queue 2 at station 2 during event C'. As $L_{22}\left(t_{F_{1}}\right) \neq 0$, the upper bound on $V_{2}$ is $L_{22}^{a}-1$. Thus,  $L_{22}\left(t_{F_{1}}\right)$ can be written as
\begin{equation}\label{eq:3.19}
L_{22}\left(t_{F_{1}}\right) = {}L_{22}^{a} - V_{2}
\end{equation}
where the conditional probability distribution of $V_{2}$ is
\begin{equation}
\displaystyle \mathop{\mathbb{P}}\Big(V_{2}={} v| V_{2} \leq L_{22}^{a}-1\Big)={}\frac{\displaystyle \mathop{\mathbb{P}}\Big(V_{2}={} v\Big)}{\sum_{v=0}^{{}L_{22}^{a}-1}\displaystyle \mathop{\mathbb{P}}\Big(V_{2}={} v\Big)}\nonumber
\end{equation}
and
\begin{equation}
\displaystyle \mathop{\mathbb{P}}\Big(V_{2}={} v\Big)=\Big(1/v!\Big)e^{-\mu_{22}\left(T_{C'}\right)}\Big(\mu_{22}\left(T_{C'}\right)\Big)^{v}\nonumber
\end{equation}
The time $T_{F_{1}}$ it takes for the server at station 2 to serve $L_{22}\left(t_{F_{1}}\right)$ customers of queue 2 is $h_{22}\left(L_{22}\left(t_{F_{1}}\right)\right)$. The probability $\displaystyle \mathop{\mathbb{P}}\Big(\text{Event $\text{F}_{1}$}\Big)$ that $L_{22}\left(t_{F_{1}}\right)$ customers are served at station 2 before queue 1 at station 1 empties with $L_{11}\left(t_{F_{1}}\right)$ customers at $t={}t_{F_{1}}$ is given by Equation $\left(\ref{eq:3.20}\right)$ and is determined using Result $\ref{Result7}$.
\begin{equation}\label{eq:3.20}
\begin{aligned}
\displaystyle \mathop{\mathbb{P}}\Big(\text{Event $\text{F}_{1}$}\Big)={}\displaystyle \mathop{\mathbb{P}}\Big(h_{22}\left(L_{22}\left(t_{F_{1}}\right)\right) < g_{11}\left(L_{11}\left(t_{F_{1}}\right)\right)\Big)
\end{aligned}
\end{equation}

In the event $\text{E}_{2}$,  the tagged customer and the $L_{12}^{'}\left(t_{E_{2}}\right)$ customers ahead of it are served at station 2 before the tagged customer exits the system.  The number of customers $L_{12}^{'}\left(t_{E_{2}}\right)$ ahead of the tagged customer at $t={}t_{E_{2}}$ is equal to $L_{11}^{'}\left(t_{C'}\right)$ as all the customers that were ahead of the tagged customer at station 1 are also ahead of it at station 2 since the server at station 2 was serving queue 2 during the events C' and $\text{F}_{1}$. The time it takes to serve the tagged customer and customers ahead of it at station 2 after the server at station 2 begins its second cycle after the arrival of the tagged customer is $h_{12}\left(L_{12}^{'}\left(t_{E_{2}}\right)+1\right)$.\\

Therefore, for sub-scenario $\left(m = {}1, n={}3\right)$, $\displaystyle\mathop{\mathbb{E}}\Big[W_{1}^{a}| m = {}1, n = {}3\Big]$ is equal to the sum of mean times for the events A', C', $\text{F}_{1}$, and $\text{E}_{2}$ and is given by Equation $\left(\ref{eq:3.21}\right)$.
\begin{equation}\label{eq:3.21}
\begin{aligned}
\displaystyle\mathop{\mathbb{E}}\Big[W_{1}^{a}| m = {}1, n = {}3\Big] = & {}\displaystyle\mathop{\mathbb{E}}\Big[h_{12}\Big(L_{12}^{a}+ K_{1}\Big)\Big]+ \displaystyle\mathop{\mathbb{E}}\Big[h_{11}\Big(L_{11}\left(t_{C'}\right)+1\Big)\Big]+ \\
& \displaystyle\mathop{\mathbb{E}}\Big[h_{22}\Big(L_{22}\left(t_{F_{1}}\right)\Big)\Big]+\displaystyle\mathop{\mathbb{E}}\Big[h_{12}\left(L_{12}^{'}\left(t_{E_{2}}\right)+1\right)\Big]
\end{aligned}
\end{equation}

The probability of this sub-scenario $\displaystyle \mathop{\mathbb{P}}\Big(n ={}3| m={}1\Big)$ is given by Equation $\left(\ref{eq:3.22}\right)$.
\begin{equation}\label{eq:3.22}
\displaystyle \mathop{\mathbb{P}}\Big(n ={}3| m={}1\Big) = \displaystyle \mathop{\mathbb{P}}\Big(\text{Event A'}\Big)\displaystyle \mathop{\mathbb{P}}\Big(\text{Event C'}\Big)\displaystyle \mathop{\mathbb{P}}\Big(\text{Event $\text{F}_{1}$}\Big)
\end{equation}

\subsection{Waiting Time Analysis of Sub-scenario$\colon \left(m = {}1, n = {}4\right)$}\label{M1N4}
\begin{figure}[h!]
\begin{center}
\includegraphics[scale=0.35]{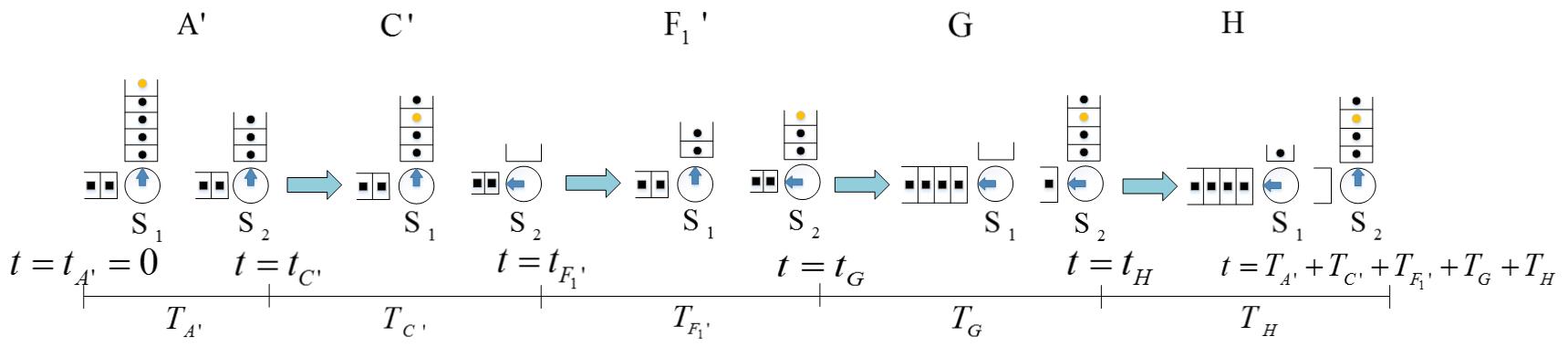}
\caption{Sequence of events for sub-scenario with $m = {}1$, $n={}4$.}
\label{fig:mesh3.14}
\end{center}
\end{figure}

From the perspective of the tagged customer arriving at station 1 at $t={}0$, the sub-scenario $\left(m = {}1, n = {}4\right)$ corresponds to a sequence of 5 events: A', C', $\text{F'}_{1}$, G, and H as shown in Figure $\ref{fig:mesh3.14}$. Events A' and C' in sub-scenario $\left(m = {}1, n = {}4\right)$ are identical to events A' and C' in sub-scenario $\left(m = {}1, n = {}3\right)$.\\

\graphicspath {{Figures/}}

In the event $\text{F'}_{1}$, queue 1 at station 1 with external arrivals empty before queue 2 at station 2. The time $T_{F'_{1}}$ to empty queue 1 at station 1 with  $L_{11}\left(t_{F'_{1}}\right)$ customers is $g_{11}\Big(L_{11}\left(t_{F'_{1}}\right)\Big)$ where $L_{11}\left(t_{F'_{1}}\right)$ is given by Equation $\left(\ref{eq:3.19}\right)$. The probability $\displaystyle \mathop{\mathbb{P}}\Big(\text{Event $\text{F'}_{1}$}\Big)$ that queue 1 at station 1 with $L_{11}\left(t_{F'_{1}}\right)$ customers empties before $L_{22}\left(t_{F'_{1}}\right)$ type 2 customers are served at station 2 is given by Equation $\left(\ref{eq:3.23}\right)$ and is determined using Result $\ref{Result7}$.
\begin{equation}\label{eq:3.23}
\begin{aligned}
\displaystyle \mathop{\mathbb{P}}\Big(\text{Event $\text{F'}_{1}$}\Big)={}\displaystyle \mathop{\mathbb{P}}\Big(g_{11}\left(L_{11}\left(t_{F'_{1}}\right)\right) < h_{22}\left(L_{22}\left(t_{F'_{1}}\right)\right)\Big)
\end{aligned}
\end{equation}

In the event G, queue 2 at station 2 with additional arrivals from station 1 empties before queue 2 at station 1 with additional external arrivals. Let $V_{3}$ be the number of type 2 customers served at station 2 during events C' and $\text{F'}_{1}$. As $L_{22}\left(t_{G}\right) \neq 0$, the upper bound on $V_{3}$ is $L_{22}\left(t_{C'}\right)-1$ for event G to occur. The queue length $L_{22}\left(t_{G}\right)$ is given by Equation $\left(\ref{eq:3.24}\right)$.
\begin{equation}\label{eq:3.24}
L_{22}\left(t_{G}\right) = {}L_{22}\left(t_{C'}\right) - V_{3}
\end{equation}
where the conditional probability distribution of $V_{3}$ is
\begin{equation}
\displaystyle \mathop{\mathbb{P}}\Big(V_{3}={} v| V_{3} \leq L_{22}\left(t_{C'}\right)-1\Big)={}\frac{\displaystyle \mathop{\mathbb{P}}\Big(V_{3}={} v\Big)}{\sum_{v=0}^{{}L_{22}\left(t_{C'}\right)-1}\displaystyle \mathop{\mathbb{P}}\Big(V_{3}={} v\Big)}\nonumber
\end{equation}
and
\begin{equation}
\displaystyle \mathop{\mathbb{P}}\Big(V_{3}={} v\Big)=\Big(1/v!\Big)e^{-\mu_{22}\left(T_{C'} + T_{F'_{1}}\right)}\Big(\mu_{22}\left(T_{C'} + T_{F'_{1}}\right)\Big)^{v}\nonumber
\end{equation}
The hitting time to 0, $T_{G}$, for queue 2 with $L_{22}\left(t_{G}\right)$ customers at station 2 and $L_{21}\left(t_{G}\right)$ type 2 customers at station 1 is $\phi_{\left(L_{21}\left(t_{G}\right), L_{22}\left(t_{G}\right)\right)}$ and is determined using Result $\ref{Result3}$ where $x={}2$ and $y={}1$. During the time $\left(0, t_{G}\right)$, additional $ S_{2}\left(t_{G}\right)$ type 2 customers arrive at station 1 where $S_{2}\left(t_{G}\right)$ is a Poisson random with parameter $\lambda_{2}t_{G}$. We write $L_{21}\left(t_{G}\right)$ as
\begin{equation}\label{eq:3.25}
L_{21}\left(t_{G}\right)={} L_{21}^{a}+ S_{2}\left(t_{G}\right)
\end{equation}

The probability $\displaystyle \mathop{\mathbb{P}}\Big(\text{Event G}\Big)$ that all existing customers of queue 2 at station 2 and additional arrivals from station 1 are served before all existing customers of queue 2 at station 1 and additional external arrivals is given by Equation $\left(\ref{eq:3.26}\right)$ and is determined using Result $\ref{Result2}$.
\begin{equation}\label{eq:3.26}
\displaystyle\mathop{\mathbb{P}}\Big(\text{Event G}\Big) ={}\sum_{k={}1}^{\infty}\alpha_{\left(k, 0\right)}\left(L_{21}\left(t_{G}\right), L_{22}\left(t_{G}\right)\right)
\end{equation}

In the event H, the tagged customer and the $L_{12}^{'}\left(t_{H}\right)$ customers ahead of it are served at station 2 before the tagged customer exits the system, where $L_{12}^{'}\left(t_{H}\right)$ is $L_{11}^{a}- K_{1}$. The time taken to serve the tagged customer and customers ahead of it at station 2 after server at station 2 begins its second cycle is $.h_{12}\left(L_{12}^{'}\left(t_{H}\right)+1\right)$.\\

For sub-scenario $\left(m = {}1, n={}4\right)$, $\displaystyle\mathop{\mathbb{E}}\Big[W_{1}^{a}| m = {}1, n = {}4\Big]$ is equal to sum of mean times for the events A', C', $\text{F'}_{1}$, G, and H and is given by Equation $\left(\ref{eq:3.27}\right)$.
\begin{equation}\label{eq:3.27}
\begin{aligned}
\displaystyle\mathop{\mathbb{E}}\Big[W_{1}^{a}| m = {}1, n = {}4\Big] = & {}\displaystyle\mathop{\mathbb{E}}\Big[h_{12}\Big(L_{12}^{a}+ K_{1}\Big)\Big]+ \displaystyle\mathop{\mathbb{E}}\Big[h_{11}\Big(L_{11}\left(t_{C'}\right)+1\Big)\Big]+ \\
& \displaystyle\mathop{\mathbb{E}}\Big[g_{11}\Big(L_{11}\left(t_{F'_{1}}\right)\Big)\Big]+ \phi_{\left(L_{21}\left(t_{G}\right), L_{22}\left(t_{G}\right)\right)} + \displaystyle\mathop{\mathbb{E}}\Big[h_{12}\left(L_{12}^{'}\left(t_{H}\right)+1\right)\Big]
\end{aligned}
\end{equation}
The probability of this sub-scenario $\displaystyle \mathop{\mathbb{P}}\Big(n ={}4| m={}1\Big)$ is given by Equation $\left(\ref{eq:3.28}\right)$.
\begin{equation}\label{eq:3.28}
\displaystyle \mathop{\mathbb{P}}\Big(n ={}4| m={}1\Big) = {}\displaystyle \mathop{\mathbb{P}}\Big(\text{Event A'}\Big)\displaystyle \mathop{\mathbb{P}}\Big(\text{Event C'}\Big)\displaystyle \mathop{\mathbb{P}}\Big(\text{Event $\text{F'}_{1}$}\Big)\displaystyle \mathop{\mathbb{P}}\Big(\text{Event G}\Big)
\end{equation}
\subsection{Waiting Time Analysis of Additional Sub-scenarios}\label{FurtherAnalysis}
As we analyze additional sub-scenarios for $m={}1$, the events begin to repeat as shown in Figure $\ref{fig:mesh3.10}$. The additional sub-scenarios $\left(m={}1, n \geq 5\right)$, all share the initial set of events A', C', and $\text{F'}_{1}$. The probability and conditional waiting times for these events were derived in the Section \ref{M1N1} -- \ref{M1N4}. In sub-scenarios $\left(m={}1, n \geq 5\right)$, event G' is immediately followed by event $\text{F}_{2}$ or $\left(\text{F'}_{2}\right)$ which is identical to event $\text{F}_{1}$ or $\left(\text{F'}_{1}\right)$ with respect to the queues being served at both the stations, but differs in terms of the queue length at the start of the event. Next, we derive the expressions for the expected waiting times for event G' and event $\text{F}_{2}$ $\left(\text{F'}_{2}\right)$.\\

In event G', queue 2 at station 1 with additional external arrival empties before queue 2 at station 2 with additional arrivals from station 1. The hitting time to 0, $T_{G'}$, for queue 2  with $L_{21}\left(t_{G'}\right)$ customers at station 1 and $L_{22}\left(t_{G'}\right)$ customers at station 2 is $g_{21}\left(L_{21}\left(t_{G'}\right)\right)$ and is determined using Result $\ref{Result1}$. The length of queues $L_{21}\left(t_{G'}\right)$ and $L_{22}\left(t_{G'}\right)$ are given by Equations $\left(\ref{eq:3.25}\right)$ and Equations $\left(\ref{eq:3.24}\right)$ respectively. The probability $\displaystyle \mathop{\mathbb{P}}\Big(\text{Event $\text{G'}$}\Big)$ that all existing customers of queue 2 at station 1 and additional external arrivals are served before all existing customers of queue 2 at station 2 and additional arrivals from station 1 is the complement of $\displaystyle \mathop{\mathbb{P}}\Big(\text{Event G'}\Big)$ in sub-scenario $\left(m = {}1, n = {}4\right)$.\\ 

The queue lengths at $t={}t_{F_{2}}$ or $t={}t_{F'_{2}}$ are given by Equation $\left(\ref{eq:29}\right)$ where the upper bound of $V_{4}$ is $L_{22}\left(t_{G'}\right)-1$.
\begin{equation}\label{eq:29}
\begin{aligned}
L_{11}\left(t_{F_{2}}\right) & ={} S_{1}\left(T_{G'}\right)\\
L_{21}\left(t_{F_{2}}\right) & ={} 0\\
L_{22}\left(t_{F_{2}}\right) & = {}L_{22}\left(t_{G'}\right)-V_{4}\\
L_{12}^{'}\left(t_{F_{2}}\right) & ={} L_{11}^{a} - K_{1}
\end{aligned}
\end{equation}

The total time for which the tagged customer has been in the system at the beginning of event $\text{F}_{2}$ $\left(\text{F'}_{2}\right)$ is equal to the sum of mean times for the events A', C', $\text{F'}_{1}$, and G'. For sub-scenarios $\left(m = {}1, n \geq{}5\right)$, $\displaystyle\mathop{\mathbb{E}}\Big[W_{1}^{a}| m = {}1, n \geq {}5\Big]$ is  given by Equation $\left(\ref{eq:3.30}\right)$.
\begin{equation}\label{eq:3.30}
\begin{aligned}
\displaystyle\mathop{\mathbb{E}}\Big[W_{1}^{a}| m = {}1, n \geq {}5\Big] = & {}\displaystyle\mathop{\mathbb{E}}\Big[h_{12}\Big(L_{12}^{a}+ K_{1}\Big)\Big]+ \displaystyle\mathop{\mathbb{E}}\Big[h_{11}\Big(L_{11}\left(t_{C'}\right)+1\Big)\Big]+ \\
& \displaystyle\mathop{\mathbb{E}}\Big[g_{11}\Big(L_{11}\left(t_{F'_{1}}\right)\Big)\Big]+ \displaystyle\mathop{\mathbb{E}}\Big[g_{21}\Big(L_{21}\left(t_{G'}\right)\Big)\Big]+  \displaystyle\mathop{\mathbb{E}}\Big[\Delta_{1}\Big]
\end{aligned}
\end{equation}

where $ \displaystyle\mathop{\mathbb{E}}\Big[\Delta_{1}\Big]$ denotes the expected waiting time for the events that follow G' till the tagged customer leaves the system. We let $\displaystyle\mathop{\mathbb{P}}\Big(\Delta_{1}\Big)$ denote the probability of $\Delta_{1}$. The probability $\displaystyle \mathop{\mathbb{P}}\Big(n \geq{}5| m={}1\Big)$ that the tagged customer is served at some time after the sequence of events A, C, $\text{F}_{1}$, and G' is given by Equation $\left(\ref{eq:3.31}\right)$.
\begin{equation}\label{eq:3.31}
\displaystyle \mathop{\mathbb{P}}\Big(n \geq{}5| m={}1\Big) = {}\displaystyle \mathop{\mathbb{P}\Big(\text{Event A'}\Big)}\displaystyle \mathop{\mathbb{P}}\Big(\text{Event C'}\Big)\displaystyle \mathop{\mathbb{P}}\Big(\text{Event $\text{F'}_{1}$}\Big)\displaystyle \mathop{\mathbb{P}}\Big(\text{Event $\text{G'}$}\Big)\displaystyle \mathop{\mathbb{P}}\Big(\Delta_{1}\Big)
\end{equation}

In the numerical computation of $\displaystyle\mathop{\mathbb{E}}\Big[W_{1}^{a}| m = {}1\Big]$ described in Section \ref{NumericalAnalysis}, if $\displaystyle \mathop{\mathbb{P}}\Big(\Delta_{1}\Big)$ is larger than the acceptable threshold, we explore additional sub-scenarios using the same approach. If $\displaystyle \mathop{\mathbb{P}}\Big(\Delta_{1}\Big)$ is less than the acceptable threshold, we set $\displaystyle\mathop{\mathbb{E}}\Big[\Delta_{1}\Big] = {}0$. The waiting time $\displaystyle \mathop{\mathbb{E}}\left[W_{1}^{a}| m = {}1\right]$ can now be calculated using Equation $\left(\ref{eq:3.7}\right)$ where we use the expected waiting times for each sub-scenarios and the probability of each sub-scenarios. We analyze the scenarios 2, 3, and 4 similarly and the details are given in Appendix \ref{AppendixB}, \ref{AppendixC}, and \ref{AppendixD} respectively.
\section{Numerical Analysis}\label{NumericalAnalysis}
In this section, we present the results of the numerical experiments designed to test the accuracy of mean conditional waiting times determined using the stochastic method developed in Section \ref{WaitingTimes}. We also compare these estimates of mean conditional waiting times with the corresponding estimates obtained using deterministic analysis.\\

\subsection{Comparison With Estimates From A Simulation Model}\label{Simulation}
In this section, we compare the mean conditional waiting times under a few different settings for different vectors $\boldsymbol{L^{a}} = {}\Big(L_{11}\left(0\right), L_{21}\left(0\right), L_{12}\left(0\right), L_{22}\left(0\right)\Big)$. In each setting, we compare the results of the conditional waiting times from the stochastic model with the corresponding estimates obtained from a simulation model. The simulation model was built using Arena software (\href{https://www.arenasimulation.com/}{www.arenasimulation.com}). In the simulation model, the stations were modeled as `process' with `seize delay release' as action and customers as `entities'. The simulation starts with the tagged customer arriving at $t={}0$ and seeing the initial queue length vector $\boldsymbol{L^{a}}$. Each simulation run stops when the tagged customer leaves the system. We record the waiting time of the tagged customer in the system. A total of 800 replications were performed and the mean waiting time across these 800 replications was computed as $\displaystyle\mathop{\mathbb{E}}\Big[W_{i_{Sim}}\Big]$. The simulation ran for approximately 5 minutes for each of the experimental settings. We define Error $\left(\Delta_{W_{i}}\right)$ as $\Bigg|\frac{\displaystyle\mathop{\mathbb{E}}\Big[W_{i_{Sim}}\Big] - \displaystyle\mathop{\mathbb{E}}\Big[W_{i_{Stoch}}\Big]}{\displaystyle\mathop{\mathbb{E}}\Big[W_{i_{Sim}}\Big]}\Bigg|$.\\

First, we analyze the performance under station and product symmetry. We set the arrival rate $\lambda_{i}$ to 1 for both the customer types at station 1 in all experiment settings. We consider two different values of service rates, i.e., $\mu_{ij} \in {}\{2.22, 2.86\}$. For each value of service rate, we consider 9 different queue length vectors $\boldsymbol{L^{a}}$. Corresponding to each $\boldsymbol{L^{a}}$, we determine the mean conditional waiting times for each scenario. These mean conditional waiting times are listed in Table \ref{T:3.3}.
{\renewcommand{\arraystretch}{1.2}
\begin{landscape}
\hspace{5cm}
\begin{table}[h!]
\centering
\caption{Mean conditional waiting times in network with product and station symmetry.}\label{T:3.3}
\begin{tabular}{|C{2cm}||C{1.2cm}|C{1.2cm}|C{1.2cm}|C{1.2cm}||C{1.2cm}|C{1.2cm}|C{1.2cm}|C{1.2cm}||C{1.4cm}|C{1.4cm}|C{1.4cm}|C{1.4cm}|}
\hline
\multicolumn{13}{|c|}{$\textbf{System parameters}\colon\lambda_{i} = {}1, \mu_{ij} = {} 2.86, \rho_{j} = {} 0.70$}\\
\hline
\multicolumn{1}{|c||}{Input} & \multicolumn{4}{c||}{\textbf{Stochastic Model}} & \multicolumn{4}{c||}{\textbf{Simulation Model}} & \multicolumn{4}{c|}{\textbf{Error} $\left(\Delta_{W_{i}}\right)$}\\
\hline
&&&&&&&&&&&&\\[-1em]
$L^{a}$ & $m={}1$ &  $m={}2$ & $m={}3$ & $m={}4$  &  $m={}1$ & $m={}2$ &  $m={}3$ & $m={}4$ &  $m={}1$ & $m={}2$ &  $m={}3$ & $m={}4$\\
\hline
(1, 1, 1, 1)	&	1.60	&	1.76	&	2.81	&	2.91	&	1.48	&	1.66	&	2.93	&	2.98	&	7.66\%	&	5.83\%	&	4.11\%	&	2.41\%	\\
(3, 3, 3, 3)	&	3.07	&	4.14	&	6.13	&	5.76	&	3.27	&	4.02	&	6.26	&	6.21	&	6.16\%	&	2.96\%	&	2.03\%	&	7.18\%	\\
(6, 6, 6, 6)	&	5.01	&	7.31	&	11.37	&	10.42	&	5.36	&	7.70	&	11.53	&	11.18	&	6.68\%	&	5.06\%	&	1.37\%	&	6.77\%	\\
(1, 1, 3, 3)	&	2.20	&	3.73	&	3.46	&	4.11	&	2.19	&	3.44	&	3.74	&	4.21	&	0.66\%	&	8.47\%	&	7.43\%	&	2.30\%	\\
(1, 1, 6, 6)	&	2.94	&	6.04	&	4.11	&	6.18	&	2.91	&	6.42	&	4.36	&	6.77	&	0.93\%	&	5.90\%	&	5.84\%	&	8.69\%	\\
(3, 3, 1, 1)	&	2.37	&	2.47	&	4.73	&	4.79	&	2.40	&	2.49	&	5.21	&	5.21	&	1.26\%	&	0.68\%	&	9.31\%	&	7.97\%	\\
(6, 6, 1, 1)	&	3.54	&	3.64	&	7.58	&	7.39	&	3.65	&	3.69	&	8.22	&	8.19	&	2.79\%	&	1.39\%	&	7.88\%	&	9.73\%	\\
(3, 6, 3, 6)	&	4.05	&	6.21	&	8.93	&	8.74	&	4.46	&	6.78	&	9.78	&	9.44	&	9.27\%	&	8.43\%	&	8.68\%	&	7.39\%	\\
(6, 3, 6, 3)	&	4.79	&	5.81	&	7.38	&	7.43	&	4.96	&	5.79	&	7.71	&	7.96	&	3.54\%	&	0.52\%	&	4.30\%	&	6.70\%	\\
\hline
\hline
\multicolumn{13}{|c|}{$\textbf{System parameters}\colon\lambda_{i} = {}1, \mu_{ij} = {} 2.22, \rho_{j} = {} 0.90$}\\
\hline
\multicolumn{1}{|c||}{Input} & \multicolumn{4}{c||}{\textbf{Stochastic Model}} & \multicolumn{4}{c||}{\textbf{Simulation Model}} & \multicolumn{4}{c|}{\textbf{Error} $\left(\Delta_{W_{i}}\right)$}\\
\hline
&&&&&&&&&&&&\\[-1em]
$L^{a}$ & $m={}1$ &  $m={}2$ & $m={}3$ & $m={}4$  &  $m={}1$ & $m={}2$ &  $m={}3$ & $m={}4$ &  $m={}1$ & $m={}2$ &  $m={}3$ & $m={}4$\\
\hline
(1, 1, 1, 1)	&	2.03	&	2.32	&	3.96	&	4.06	&	1.94	&	2.19	&	4.15	&	4.21	&	4.85\%	&	5.74\%	&	4.60\%	&	3.72\%	\\
(3, 3, 3, 3)	&	3.90	&	5.30	&	8.73	&	7.97	&	4.22	&	5.25	&	8.89	&	8.74	&	7.43\%	&	1.07\%	&	1.76\%	&	8.82\%	\\
(6, 6, 6, 6)	&	6.60	&	9.40	&	16.07	&	15.26	&	7.08	&	10.06	&	17.02	&	15.96	&	6.84\%	&	6.63\%	&	5.60\%	&	4.41\%	\\
(1, 1, 3, 3)	&	2.85	&	4.94	&	4.89	&	5.67	&	2.83	&	4.65	&	5.36	&	5.96	&	0.60\%	&	6.23\%	&	8.70\%	&	4.97\%	\\
(1, 1, 6, 6)	&	3.80	&	8.17	&	5.79	&	8.55	&	3.82	&	8.80	&	6.28	&	9.39	&	0.51\%	&	7.17\%	&	7.82\%	&	8.92\%	\\
(3, 3, 1, 1)	&	3.04	&	3.18	&	6.90	&	6.78	&	3.10	&	3.23	&	7.36	&	7.32	&	1.98\%	&	1.45\%	&	6.22\%	&	7.41\%	\\
(6, 6, 1, 1)	&	4.56	&	4.69	&	10.61	&	10.61	&	4.62	&	4.68	&	11.68	&	11.62	&	1.23\%	&	0.22\%	&	9.18\%	&	8.71\%	\\
(3, 6, 3, 6)	&	5.57	&	8.50	&	13.61	&	12.66	&	5.95	&	9.32	&	14.42	&	13.75	&	6.39\%	&	8.78\%	&	5.65\%	&	7.88\%	\\
(6, 3, 6, 3)	&	6.36	&	7.47	&	10.37	&	10.03	&	6.40	&	7.42	&	10.95	&	11.04	&	0.63\%	&	0.70\%	&	5.29\%	&	9.16\%	\\
\hline
\end{tabular}
\end{table}
\end{landscape}
\normalsize

Table \ref{T:3.3} summarizes the performance of the approach developed in the present paper showing the errors for each of the scenario. Overall, the average error is 5\% while the error for 80\% of the cases is less than 9\%. One possible reason for relatively high errors in some cases is the effect of assuming queue lengths to be integers while using the results 1 -- 7 for the analysis. However, these errors appear to be acceptable in view of the complexity of the system under consideration. We observe, that in general, the conditional waiting times are lower for scenario $m={}1$, and is higher for scenarios $m={}3$ and $m={}4 $. Thus, the arriving customer waits for less time, if on arrival, the same customer type is being processed at both the stations. We also observe that the waiting times are relatively high when the queue lengths are large at station 1 than at station 2.\\

In the second set, we compute the mean conditional waiting time of the arriving customer under the settings of station asymmetry. For this, we consider two types of station asymmetry. In the first type, we choose the system parameters so that the upstream station is the bottleneck. To do so, we set the service rates $\mu_{i1}$ to 2.22, and the service rates $\mu_{i2}$ to 2.86. Thus, the traffic intensity at station 1 is $\rho_{1}={}0.90$ and at station 2 is $\rho_{2}={}0.70$. In the second type of station asymmetry, we choose the system parameter so that downstream station is the bottleneck. To study performance under the settings where the downstream station is the bottleneck, we set the service rates $\mu_{i1}$ to 2.86, and the service rates $\mu_{i2}$ to 2.22. Thus, the traffic intensity at station 1 is $\rho_{1}={}0.70$ and at station 2 is $\rho_{2}={}0.90$. The results of this study are listed in Table \ref{T:3.4}.
{\renewcommand{\arraystretch}{1.2}
\begin{landscape}
\hspace{5cm}
\begin{table}[h!]
\centering
\caption{Mean conditional waiting times in network with station asymmetry.}\label{T:3.4}
\begin{tabular}{|C{2cm}||C{1.2cm}|C{1.2cm}|C{1.2cm}|C{1.2cm}||C{1.2cm}|C{1.2cm}|C{1.2cm}|C{1.2cm}||C{1.4cm}|C{1.4cm}|C{1.4cm}|C{1.4cm}|}
\hline
\multicolumn{13}{|c|}{$\textbf{System parameters}\colon\lambda_{i} = {}1, \mu_{i1} = {} 2.22, \mu_{i2} = {} 2.86, \rho_{1} = {} 0.90, \rho_{2} = {} 0.70$}\\
\hline
\multicolumn{1}{|c||}{Input} & \multicolumn{4}{c||}{\textbf{Stochastic Model}} & \multicolumn{4}{c||}{\textbf{Simulation Model}} & \multicolumn{4}{c|}{\textbf{Error} $\left(\Delta_{W_{i}}\right)$}\\
\hline
&&&&&&&&&&&&\\[-1em]
$L^{a}$ & $m={}1$ &  $m={}2$ & $m={}3$ & $m={}4$  &  $m={}1$ & $m={}2$ &  $m={}3$ & $m={}4$ &  $m={}1$ & $m={}2$ &  $m={}3$ & $m={}4$\\
\hline
(1, 1, 1, 1)	&	1.74	&	1.82	&	3.25	&	3.34	&	1.66	&	1.74	&	3.49	&	3.51	&	4.68\%	&	4.46\%	&	6.95\%	&	4.95\%	\\
(3, 3, 3, 3)	&	3.26	&	3.98	&	6.59	&	6.35	&	3.24	&	3.83	&	6.96	&	6.93	&	0.68\%	&	3.97\%	&	5.33\%	&	8.35\%	\\
(6, 6, 6, 6)	&	5.40	&	7.06	&	11.67	&	11.04	&	5.84	&	6.99	&	12.41	&	11.95	&	7.48\%	&	1.00\%	&	6.00\%	&	7.57\%	\\
(1, 1, 3, 3)	&	2.37	&	3.48	&	4.01	&	4.35	&	2.20	&	3.23	&	4.17	&	4.46	&	7.53\%	&	7.82\%	&	3.77\%	&	2.37\%	\\
(1, 1, 6, 6)	&	3.08	&	6.31	&	4.65	&	6.45	&	2.98	&	6.07	&	5.12	&	6.72	&	3.27\%	&	3.95\%	&	9.13\%	&	3.99\%	\\
(3, 3, 1, 1)	&	2.59	&	2.68	&	5.68	&	5.67	&	2.66	&	2.68	&	6.24	&	6.23	&	2.71\%	&	0.15\%	&	9.00\%	&	9.03\%	\\
(6, 6, 1, 1)	&	4.00	&	4.10	&	9.27	&	9.48	&	4.09	&	4.13	&	10.28	&	10.27	&	2.20\%	&	0.70\%	&	9.82\%	&	7.70\%	\\
(3, 6, 3, 6)	&	4.14	&	5.53	&	9.68	&	9.68	&	4.31	&	5.67	&	10.68	&	10.39	&	4.04\%	&	2.52\%	&	9.40\%	&	6.80\%	\\
(6, 3, 6, 3)	&	5.04	&	5.78	&	8.38	&	8.02	&	5.16	&	5.70	&	8.70	&	8.67	&	2.41\%	&	1.46\%	&	3.70\%	&	7.52\%	\\
\hline
\hline
\multicolumn{13}{|c|}{$\textbf{System parameters}\colon\lambda_{i} = {}1, \mu_{i1} = {} 2.86, \mu_{i2} = {} 2.22, \rho_{1} = {} 0.70, \rho_{2} = {} 0.90$}\\
\hline
\multicolumn{1}{|c||}{Input} & \multicolumn{4}{c||}{\textbf{Stochastic Model}} & \multicolumn{4}{c||}{\textbf{Simulation Model}} & \multicolumn{4}{c|}{\textbf{Error} $\left(\Delta_{W_{i}}\right)$}\\
\hline
&&&&&&&&&&&&\\[-1em]
$L^{a}$ & $m={}1$ &  $m={}2$ & $m={}3$ & $m={}4$  &  $m={}1$ & $m={}2$ &  $m={}3$ & $m={}4$ &  $m={}1$ & $m={}2$ &  $m={}3$ & $m={}4$\\
\hline
(1, 1, 1, 1)	&	2.00	&	2.40	&	3.46	&	3.65	&	1.89	&	2.22	&	3.80	&	3.86	&	5.61\%	&	8.48\%	&	8.94\%	&	5.45\%	\\
(3, 3, 3, 3)	&	3.69	&	5.64	&	7.65	&	7.67	&	3.92	&	5.78	&	8.40	&	8.43	&	5.98\%	&	2.33\%	&	8.97\%	&	9.00\%	\\
(6, 6, 6, 6)	&	6.11	&	10.96	&	14.42	&	14.22	&	6.19	&	11.45	&	15.46	&	15.64	&	1.23\%	&	4.28\%	&	6.71\%	&	9.12\%	\\
(1, 1, 3, 3)	&	2.66	&	4.95	&	4.34	&	5.45	&	2.64	&	5.13	&	4.63	&	5.92	&	0.92\%	&	3.59\%	&	6.29\%	&	7.90\%	\\
(1, 1, 6, 6)	&	3.69	&	8.65	&	4.65	&	8.79	&	3.69	&	9.42	&	5.04	&	9.66	&	0.04\%	&	8.24\%	&	7.71\%	&	8.95\%	\\
(3, 3, 1, 1)	&	2.87	&	3.09	&	6.03	&	5.92	&	2.92	&	3.05	&	6.62	&	6.52	&	1.79\%	&	1.60\%	&	8.91\%	&	9.20\%	\\
(6, 6, 1, 1)	&	4.23	&	4.38	&	9.49	&	9.34	&	4.34	&	4.45	&	10.32	&	10.21	&	2.57\%	&	1.61\%	&	8.11\%	&	8.55\%	\\
(3, 6, 3, 6)	&	4.89	&	9.87	&	12.82	&	12.46	&	5.37	&	10.84	&	13.91	&	13.33	&	8.87\%	&	8.96\%	&	7.87\%	&	6.55\%	\\
(6, 3, 6, 3)	&	5.98	&	7.65	&	9.00	&	9.84	&	5.96	&	7.72	&	9.67	&	10.62	&	0.38\%	&	0.90\%	&	6.92\%	&	7.35\%	\\
\hline
\end{tabular}
\end{table}
\end{landscape}
\normalsize

Table \ref{T:3.4} summarizes the performance of the approach for settings with station asymmetry. Overall, the average error is 4\% while the error for 80\% of the cases is less than 9\%. In this setting, we observe that the conditional waiting times are lower when the bottleneck is at upstream as opposed to downstream. Similar to network with the symmetric stations, we observe that the conditional waiting times in this system are also lower for scenario $m={}1$.

\subsection{Comparison With Estimates From Deterministic Analysis}\label{Deterministic}
Next, we compare the results of the stochastic model with deterministic model to quantify how variability in arrival and service times affect the conditional waiting times. In the deterministic model, we assume that the arrival rate and service rates are constant with the values listed in Tables \ref{T:3.5} and Table \ref{T:3.6}. Note that in case of deterministic model, depending on the parameter setting, an arriving customer experiences only one of sub-scenarios and therefore, the probability of other sub-scenarios is zero. The comparison of results for the system with product and station symmetry are listed in Table \ref{T:3.5} and for systems with station asymmetry are listed in Table \ref{T:3.6}. We observe that because of the variability in the process, the conditional waiting times are higher by 21\% on an average.\\

We also determine the mean waiting times of the customers in the system using simulation and report it in the table. It should be noted that the conditional waiting times are dependent on the scenario that the tagged customer sees on its arrival while the mean waiting times is the unconditional waiting times that an arriving customer sees in steady state. In Table \ref{T:3.5}, we observe that the mean waiting time is 2.33 and 8.32 for traffic intensity of 0.70 and 0.90 respectively. The mean conditional waiting times varies between 1.60 to 11.37 for different values of $\boldsymbol{L^{a}}$ for traffic intensity of 0.70 while it varies between 2.03 to 16.07 for different values of $\boldsymbol{L^{a}}$ for traffic intensity of 0.90. We note that in Table \ref{T:3.6}, the mean waiting time is 5.32 irrespective of the location of the bottleneck station. The mean conditional waiting times varies between 1.74 to 11.67 for different values of $\boldsymbol{L^{a}}$ when the upstream station is the bottleneck while it varies between 2.00 to 14.42 for different values of $\boldsymbol{L^{a}}$ when the downstream station is the bottleneck.  We notice that the conditional waiting times provide more precise information that an arriving customer is likely to experience given the queue lengths and customer types being served at both the stations.\\

{\renewcommand{\arraystretch}{1.15}
\begin{table}[h!]
\centering
\caption{Comparison of conditional stochastic and deterministic waiting times.}\label{T:3.5}
\begin{tabular}{|C{2cm}||C{1.2cm}|C{1.2cm}|C{1.2cm}|C{1.2cm}||C{1.2cm}|C{1.2cm}|C{1.2cm}|C{1.2cm}|}
\hline
\multicolumn{9}{|c|}{$\textbf{System parameters}\colon\lambda_{i} = {}1, \mu_{ij} = {} 2.86, \rho_{j} = {} 0.70$}\\

\hline
\multicolumn{1}{|c||}{Input} & \multicolumn{4}{c||}{\textbf{Stochastic Model}} & \multicolumn{4}{c|}{\textbf{Deterministic model}} \\
\hline
&&&&&&&&\\[-1em]
$L^{a}$ & $m={}1$ &  $m={}2$ & $m={}3$ & $m={}4$  &  $m={}1$ & $m={}2$ &  $m={}3$ & $m={}4$\\
\hline
(1, 1, 1, 1)	&	1.60	&	1.76	&	2.81	&	2.91	&	1.05	&	1.40	&	1.75	&	1.75	\\
(3, 3, 3, 3)	&	3.07	&	4.14	&	6.13	&	5.76	&	2.45	&	3.50	&	5.60	&	4.90	\\
(6, 6, 6, 6)	&	5.01	&	7.31	&	11.37	&	10.42	&	4.55	&	6.65	&	11.53	&	9.45	\\
(1, 1, 3, 3)	&	2.20	&	3.73	&	3.46	&	4.11	&	1.75	&	3.50	&	1.75	&	3.50	\\
(1, 1, 6, 6)	&	2.94	&	6.04	&	4.11	&	6.18	&	2.80	&	5.95	&	2.80	&	5.95	\\
(3, 3, 1, 1)	&	2.37	&	2.47	&	4.73	&	4.79	&	1.75	&	2.10	&	3.50	&	3.50	\\
(6, 6, 1, 1)	&	3.54	&	3.64	&	7.58	&	7.39	&	2.80	&	3.15	&	5.95	&	5.95	\\
(3, 6, 3, 6)	&	4.05	&	6.21	&	8.93	&	8.74	&	2.45	&	8.40	&	8.75	&	8.40	\\
(6, 3, 6, 3)	&	4.79	&	5.81	&	7.38	&	7.43	&	4.55	&	5.60	&	4.55	&	7.00	\\
\hline
\multicolumn{9}{|c|}{$\textbf{Mean waiting times}\colon W_{i} = {}2.33$; $\textbf{Range of } \displaystyle\mathop{\mathbb{E}}\Big[W_{1}^{a}| m\Big]\colon\left[1.60, 11.37\right]$}\\
\hline
\hline
\multicolumn{9}{|c|}{$\textbf{System parameters}\colon\lambda_{i} = {}1, \mu_{ij} = {} 2.22, \rho_{j} = {} 0.90$}\\
\hline
\multicolumn{1}{|c||}{Input} & \multicolumn{4}{c||}{\textbf{Stochastic Model}} & \multicolumn{4}{c|}{\textbf{Deterministic model}}\\
\hline
&&&&&&&&\\[-1em]
$L^{a}$ & $m={}1$ &  $m={}2$ & $m={}3$ & $m={}4$  &  $m={}1$ & $m={}2$ &  $m={}3$ & $m={}4$\\
\hline
(1, 1, 1, 1)	&	2.03	&	2.32	&	3.96	&	4.06	&	1.35	&	1.80	&	2.25	&	2.25	\\
(3, 3, 3, 3)	&	3.90	&	5.30	&	8.73	&	7.97	&	3.15	&	4.50	&	8.10	&	6.30	\\
(6, 6, 6, 6)	&	6.60	&	9.40	&	16.07	&	15.26	&	5.85	&	8.55	&	17.11	&	13.06	\\
(1, 1, 3, 3)	&	2.85	&	4.94	&	4.89	&	5.67	&	2.25	&	4.50	&	2.25	&	4.50	\\
(1, 1, 6, 6)	&	3.80	&	8.17	&	5.79	&	8.55	&	3.60	&	8.55	&	3.60	&	8.55	\\
(3, 3, 1, 1)	&	3.04	&	3.18	&	6.90	&	6.78	&	2.25	&	2.70	&	4.50	&	4.50	\\
(6, 6, 1, 1)	&	4.56	&	4.69	&	10.61	&	10.61	&	3.60	&	4.05	&	8.55	&	8.55	\\
(3, 6, 3, 6)	&	5.57	&	8.50	&	13.61	&	12.66	&	3.15	&	12.16	&	13.06	&	12.16	\\
(6, 3, 6, 3)	&	6.36	&	7.47	&	10.37	&	10.03	&	5.85	&	7.20	&	5.85	&	9.00	\\
\hline
\multicolumn{9}{|c|}{$\textbf{Mean waiting times}\colon W_{i} = {}8.32$; $\textbf{Range of } \displaystyle\mathop{\mathbb{E}}\Big[W_{1}^{a}| m\Big]\colon\left[2.03, 16.07\right]$}\\
\hline
\end{tabular}
\end{table}
\normalsize

{\renewcommand{\arraystretch}{1.15}
\begin{table}[h!]
\centering
\caption{Comparison of conditional stochastic and deterministic waiting times.}\label{T:3.6}
\begin{tabular}{|C{2cm}||C{1.2cm}|C{1.2cm}|C{1.2cm}|C{1.2cm}||C{1.2cm}|C{1.2cm}|C{1.2cm}|C{1.2cm}|}
\hline
\multicolumn{9}{|c|}{$\textbf{System parameters}\colon\lambda_{i} = {}1, \mu_{i1} = {} 2.22, \mu_{i2} = {} 2.86, \rho_{1} = {} 0.90, \rho_{2} = {} 0.70$}\\
\hline
\multicolumn{1}{|c||}{Input} & \multicolumn{4}{c||}{\textbf{Stochastic Model}} & \multicolumn{4}{c|}{\textbf{Deterministic model}} \\
\hline
&&&&&&&&\\[-1em]
$L^{a}$ & $m={}1$ &  $m={}2$ & $m={}3$ & $m={}4$  &  $m={}1$ & $m={}2$ &  $m={}3$ & $m={}4$\\
\hline
(1, 1, 1, 1)	&	1.74	&	1.82	&	3.25	&	3.34	&	1.40	&	1.40	&	1.75	&	1.75	\\
(3, 3, 3, 3)	&	3.26	&	3.98	&	6.59	&	6.35	&	2.45	&	3.50	&	4.90	&	4.90	\\
(6, 6, 6, 6)	&	5.40	&	7.06	&	11.67	&	11.04	&	4.55	&	6.65	&	10.15	&	10.15	\\
(1, 1, 3, 3)	&	2.37	&	3.48	&	4.01	&	4.35	&	1.75	&	2.80	&	1.75	&	3.15	\\
(1, 1, 6, 6)	&	3.08	&	6.31	&	4.65	&	6.45	&	2.80	&	5.95	&	2.80	&	5.95	\\
(3, 3, 1, 1)	&	2.59	&	2.68	&	5.68	&	5.67	&	2.15	&	2.15	&	3.95	&	3.95	\\
(6, 6, 1, 1)	&	4.00	&	4.10	&	9.23	&	9.48	&	3.50	&	3.50	&	8.00	&	8.00	\\
(3, 6, 3, 6)	&	4.14	&	5.53	&	9.68	&	9.68	&	2.45	&	4.55	&	8.05	&	8.05	\\
(6, 3, 6, 3)	&	5.04	&	5.78	&	8.38	&	8.02	&	4.55	&	5.60	&	7.00	&	7.00	\\
\hline
\multicolumn{9}{|c|}{$\textbf{Mean waiting times}\colon W_{i} = {}5.32$; $\textbf{Range of } \displaystyle\mathop{\mathbb{E}}\Big[W_{1}^{a}| m\Big]\colon\left[1.74, 11.67\right]$}\\
\hline
\hline
\multicolumn{9}{|c|}{$\textbf{System parameters}\colon\lambda_{i} = {}1, \mu_{i1} = {} 2.86, \mu_{i2} = {} 2.22, \rho_{1} = {} 0.70, \rho_{2} = {} 0.90$}\\
\hline
\multicolumn{1}{|c||}{Input} & \multicolumn{4}{c||}{\textbf{Stochastic Model}} & \multicolumn{4}{c|}{\textbf{Deterministic model}} \\
\hline
&&&&&&&&\\[-1em]
$L^{a}$ & $m={}1$ &  $m={}2$ & $m={}3$ & $m={}4$  &  $m={}1$ & $m={}2$ &  $m={}3$ & $m={}4$\\
\hline
(1, 1, 1, 1)	&	2.00	&	2.40	&	3.46	&	3.65	&	1.35	&	1.80	&	2.25	&	2.25	\\
(3, 3, 3, 3)	&	3.69	&	5.64	&	7.65	&	7.67	&	3.15	&	4.50	&	8.10	&	7.20	\\
(6, 6, 6, 6)	&	6.11	&	10.96	&	14.42	&	14.22	&	5.85	&	8.55	&	14.11	&	14.86	\\
(1, 1, 3, 3)	&	2.66	&	4.95	&	4.34	&	5.45	&	2.25	&	4.50	&	2.25	&	4.50	\\
(1, 1, 6, 6)	&	3.69	&	8.65	&	4.65	&	8.79	&	3.60	&	8.55	&	3.60	&	8.55	\\
(3, 3, 1, 1)	&	2.87	&	3.09	&	6.03	&	5.92	&	2.25	&	2.70	&	4.50	&	4.50	\\
(6, 6, 1, 1)	&	4.23	&	4.38	&	9.49	&	9.34	&	3.60	&	4.05	&	7.65	&	7.65	\\
(3, 6, 3, 6)	&	4.89	&	9.87	&	12.82	&	12.46	&	3.15	&	10.16	&	13.03	&	12.16	\\
(6, 3, 6, 3)	&	5.98	&	7.65	&	9.00	&	9.84	&	5.85	&	7.20	&	5.85	&	9.00	\\
\hline
\multicolumn{9}{|c|}{$\textbf{Mean waiting times}\colon W_{i} = {}5.32$; $\textbf{Range of } \displaystyle\mathop{\mathbb{E}}\Big[W_{1}^{a}| m\Big]\colon\left[2.00, 14.42\right]$}\\
\hline
\end{tabular}
\end{table}
\normalsize
\section{Extensions to More General Settings}\label{Extensions}
In this section, we comment on some extensions of this analysis. We discuss the extension of our approach in two specific directions, namely, analysis of systems with switchover times, and analysis of networks with more than two stations. In each case, the exact analysis is computationally prohibitive, and so we discuss possible approaches based on approximations.\\

$\textbf{Extension to systems with switchover times}\colon$ Exact analysis of systems with switchover times is complex. However, with minor modifications, we can use the approach described in Section \ref{WaitingTimes} to analyze the system described in Figure \ref{fig:mesh3.1} with switchover times. Consider the system described in Figure \ref{fig:mesh3.1} but with non-zero-switchover times, i.e., a switchover time $H_{ij}$ is required when the server switches from queue $i-1$ to queue $i$ at station $j$. We assume that these switchover times are identically distributed exponential random variables, independent of any other events involved $\left(\text{e.g., service times, other queues, etc.}\right)$. Let the first moment of the switchover time be $\mu_{H_{ij}}^{-1}$. We assume that the switchover times are state independent, i.e., the server incurs a switchover time at the polled queue whether or not customers are waiting at the queue. Without loss of generality, we also assume that when the tagged customer arrives at station 1, there is no setup of servers at that instant at both the stations.\\

To determine the mean conditional waiting times for a particular sub-scenario, we add the switchover times for all the server switchovers in the conditional waiting times of the corresponding sub-scenario. However, determining the probability of occurrence of a sub-scenario requires additional results in which the switchover times are considered in addition to the service times of the customers queued at the stations. The waiting times for different sub-scenarios for scenario $m={}1$ for the described system are summarized in Table \ref{T:3.7}. We provide details for one sub-scenario, namely, $n={}2$ for illustration purpose.\\

\renewcommand{\arraystretch}{1.80}
\begin{table}[h]
\centering
\caption{Waiting times in different subscenarios for scenario $m={}1$ for non-zero-switchover times.}\label{T:3.7}
\begin{tabular}{ |c|c|p{11cm}| } 
\hline
Scenario & Sub-scenario & Expected Waiting Time \\ 
\hline
$m={}1$ & $n = {}1   $ &  $\displaystyle\mathop{\mathbb{E}}\Big[h_{12}\Big(L_{11}^{a} + L_{12}^{a}+ 1\Big)\Big]$\\ 
$m={}1$ & $n = {}2   $ &  $\displaystyle\mathop{\mathbb{E}}\Big[h_{12}\Big(L_{12}^{a}+ K_{1}\Big) + h_{22}\Big(L_{22}\left(t_{c}\right)\Big) + W'_{L_{11}^{'}\left(t_{D}\right), L_{12}\left(t_{D}\right)}\Big] + \mu_{H_{22}}^{-1}$\\ 
$m={}1$ & $n = {}3   $ &  $\displaystyle\mathop{\mathbb{E}}\Big[h_{12}\Big(L_{12}^{a}+ K_{1}\Big) + h_{11}\Big(L_{11}\left(t_{C'}\right)+1\Big) + h_{22}\Big(L_{22}\left(t_{F_{1}}\right)\Big) + h_{12}\left(L_{12}^{'}\left(t_{E_{2}}\right)+1\right)\Big]+ \mu_{H_{12}}^{-1}$\\
$m={}1$ & $n = {}4   $ &  $\displaystyle\mathop{\mathbb{E}}\Big[h_{12}\Big(L_{12}^{a}+ K_{1}\Big) + h_{11}\Big(L_{11}\left(t_{C'}\right)+1\Big) + g_{11}\Big(L_{11}\left(t_{F'_{1}}\right)\Big) + h_{12}\left(L_{12}^{'}\left(t_{H}\right)+1\right)\Big] + \phi_{\left(L_{21}\left(t_{G}\right), L_{22}\left(t_{G}\right)\right)}+ \mu_{H_{12}}^{-1}$\\
\hline
\end{tabular}
\end{table}
\normalsize

Recall that this sub-scenario involves events A', C, D, and $\text{E}_{1}$. In event A', exactly $K_{1}$ customers, where $K_{1} \leq L_{11}^{a}$, are served at station 1 before station 2 server switches to serve queue 2  at $t={}t_{C}$ after serving $ L_{12}^{a}+ K_{1}$ type 1 customers. Since no switchover time is incurred at this time $\left(\text{assumption}\right)$, the time $T_{A}$ it takes for server at station 2 to serve $ L_{12}^{a}+ K_{1}$ customers of queue 1 is $h_{12}\Big(L_{12}^{a}+ K_{1}\Big)$. In event C, the switchover at station 2 for type 2 customers and the service of all customers of queue 2 at station 2 at $t={}t_{C}$ happens before the tagged customer is served at station 1. The time $T_{C}$ it takes for the server at station 2 to perform switchover and serve $L_{22}\left(t_{C}\right)$ customers of queue 2 is $\mu_{H_{22}}^{-1} + h_{22}\left(L_{22}\left(t_{C}\right)\right)$. In the events D and $\text{E}_{1}$, the tagged customer first gets served at station 1 after the service of $L_{11}^{'}\left(t_{D}\right)$ customers ahead of it and then moves to station 2 at $t={}t_{E_{1}}$, where it eventually gets served and exits the system. The duration of events D and $\text{E}_{1}$ and can be obtained by modifying Result \ref{Result5} to incorporate switchover times. Let  $\displaystyle\mathop{\mathbb{E}}\Big[W'_{L_{11}^{'}\left(t_{D}\right), L_{12}\left(t_{D}\right)}\Big]$ be the expected duration of events D and $\text{E}_{1}$. For sub-scenario $\left(m = {}1, n={}2\right)$, $\displaystyle\mathop{\mathbb{E}}\Big[W_{1}^{a}| m = {}1, n = {}2\Big]$ is equal to expected sum of duration of events A', C, D, and $\text{E}_{1}$ and is given by Equation $\left(\ref{eq:3.32}\right)$.
\begin{equation}\label{eq:3.32}
\begin{aligned}
\displaystyle\mathop{\mathbb{E}}\Big[W_{1}^{a}| m = {}1, n = {}2\Big] =  {}\displaystyle\mathop{\mathbb{E}}\Big[h_{12}\Big(L_{12}^{a}+ K_{1}\Big) + h_{22}\Big(L_{22}\left(t_{c}\right)\Big) + W'_{L_{11}^{'}\left(t_{D}\right), L_{12}\left(t_{D}\right)}\Big] + \mu_{H_{22}}^{-1}
\end{aligned}
\end{equation}

The probability $\displaystyle \mathop{\mathbb{P}}\Big(\text{Event A'}\Big)$ is the complement of $\displaystyle \mathop{\mathbb{P}}\Big(n ={}1| m={}1\Big)$. The probability $\displaystyle \mathop{\mathbb{P}}\Big(\text{Event C}\Big)$ that the setup at station 2 for type 2 customers and the service of  $L_{22}\left(t_{C}\right)$ customers of queue 2 at station 2 happens before $L_{11}^{'}\left(t_{C}\right)+1$ customer gets served at station 1 is given by Equation $\left(\ref{eq:3.33}\right)$.
\begin{equation}\label{eq:3.33}
\begin{aligned}
\displaystyle \mathop{\mathbb{P}}\Big(\text{Event C}\Big) ={}\displaystyle \mathop{\mathbb{P}}\Big(h_{22}\Big(L_{22}\left(t_{C}\right)\Big) + H_{22}< {} h_{11}\left(L_{11}^{'}\left(t_{C}\right)+1\right)\Big)
\end{aligned}
\end{equation}

The probability of this sub-scenario $\displaystyle \mathop{\mathbb{P}}\Big(n ={}2| m={}1\Big)$ that the tagged customer is served after the sequence of events A', C, D, and $\text{E}_{1}$ is given by Equation $\left(\ref{eq:3.34}\right)$.
\begin{equation}\label{eq:3.34}
\displaystyle \mathop{\mathbb{P}}\Big(n ={}2| m={}1\Big) = \displaystyle \mathop{\mathbb{P}}\Big(\text{Event A'}\Big)\displaystyle \mathop{\mathbb{P}}\Big(\text{Event C}\Big)
\end{equation}

Using a similar modification to the proposed approach, we can analyze other sub-scenarios of scenario $m={}1$ and the remaining scenarios for the case of non-zero-switchover times.\\

$\textbf{Extension to \emph{n} stations}\colon$ To analyze a larger network of tandem polling queues with \emph{n} stations, where $n > 2$, we propose an approximate method that analyzes two stations at a time. In this approach, we analyze the network in groups of two consecutive stations. For each group, the analysis done in this paper can be used to estimate conditional waiting times. Another approach is to only analyze the group of two stations that contain the bottleneck station; especially, if there is a strong bottleneck present in the system. Once the bottleneck station is determined, the conditional waiting times for the bottleneck stations can be determined using approach described in this paper.
\section{Conclusions}\label{Conclusions}
In this paper, we derive mean waiting times in a tandem network of polling queues conditioned on the state of the system/network seen by the arriving customer. We believe that these conditional mean waiting times are of significant importance to providing better and more realistic estimates of waiting times a customer is likely to see in the network. We use a sample path analysis approach that classifies the system state seen upon arrival into four possible scenarios. In the analysis of each scenario, we leverage patterns of repeating sequence of events that occur before the customer leaves the network. These patterns, significantly reduce the complexity of our sample path analysis; allowing the estimation to be done using a simple numerical algorithm procedure. We show that our approach provides reasonably accurate estimates of mean conditional waiting times. We also show that these waiting times can be quite different from mean waiting times seen by an arbitrary customer. We also provide discussion on how our work can be extended to estimate conditional waiting times in more general systems in the future.

\pagebreak
\appendix
\section{Appendix A}\label{AppendixA}
\textbf{Result \ref{Result2}}. Let $p_{1}\left(u, w\right)$ be the probability that station 1 empties before station 2 and let $p_{2}\left(u, w\right)$ be the probability that station 2 empties before station 1. Then, for $k > 0\colon$
\begin{align*}
p_{1}\left(u, w\right) & = {}1 - \sum_{k={}1}^{\infty}\alpha_{\left(k, 0\right)}\left(u, w\right)\\
p_{2}\left(u, w\right) & = {}\sum_{k={}1}^{\infty}\alpha_{\left(k, 0\right)}\left(u, w\right)
\end{align*}

\begin{proof}
For station 2 to empty before station 1 starting from state $\left(u, w\right)$, the system must transition to a state $\left(k, 0\right)$ for $k > 0$, before it transitions to a state $\left(0, k'\right)$ for $k' > 0$. For station 1 to empty before station 2, the system must transition to a state $\left(0, k\right)$ for $k > 0$, before it transitions to state $\left(k', 0\right)$ for $k' > 0$. Considering the transitions of the Markov process, the required probability can be determined by assuming states of form $\left(0, k\right)$ and $\left(k, 0\right)$ as absorbing states and all other states of the form $\left(i, j\right)$ as irreducible transient states where $i$, $j$, and $k$ are positive integers. Define $\mathcal{R}_{1} = {} \Big\{\, \left(0, k\right) \mid k \in \displaystyle\mathop{\mathbb{Z^{+}}}\,\Big\}$ and $\mathcal{R}_{2} = {} \Big\{\,\left(k, 0\right) \mid k \in \displaystyle\mathop{\mathbb{Z^{+}}}\,\Big\}$ as two disjoint sets of recurrent classes and $\mathcal{T} = {} \Big\{\,\left(i, j\right) \mid i, j \in\displaystyle\mathop{\mathbb{Z^{+}}}\,\Big\}$ as set of transient classes such that $\mathcal{S} = {} \mathcal{R}_{1} \cup \mathcal{R}_{2} \cup \mathcal{T}$. Let $\left\{\,X_{n}, n\geq 0\,\right\}$ be the embedded Markov chain for $X\left(t\right)$ where the transition from one state to another is according to the one-step transition probabilities $P_{\left(ij\right), \left(i'j'\right)} = {}\left\{\,X_{n+1} = \left(i', j'\right)| X_{n} = \left(i, j\right)\,\right\}$. We write the transition matrix $P$ of the embedded Markov chain in the form of fundamental matrix after possible reordering of the states as$\colon$

\[P = {}
\renewcommand\arraystretch{0.8}
\mleft[
\begin{array}{c|c}
I    &    O\\
\hline
S    &    Q\\
\end{array}
\mright]
\]
where sub-matrices $I_{\left\{\mathcal{R}_{1} \cup \mathcal{R}_{2}\right\} \times \left\{\mathcal{R}_{1} \cup \mathcal{R}_{2}\right\}}$, $O_{\left\{\mathcal{R}_{1} \cup \mathcal{R}_{2}\right\} \times \mathcal{T}}$, $S_{\mathcal{T} \times \left\{\mathcal{R}_{1} \cup \mathcal{R}_{2}\right\}}$, and $Q_{\mathcal{T} \times \mathcal{T}}$ represents the respective transition probabilities from $\left\{\mathcal{R}_{1} \cup \mathcal{R}_{2}\right\}$ to $\left\{\mathcal{R}_{1} \cup \mathcal{R}_{2}\right\}$, $\left\{\mathcal{R}_{1} \cup \mathcal{R}_{2}\right\}$ to $\mathcal{T}$, $\mathcal{T}$ to $\left\{\mathcal{R}_{1} \cup \mathcal{R}_{2}\right\}$, and $\mathcal{T}$ to $\mathcal{T}$. Let $I'_{\mathcal{T} \times \mathcal{T}}$ be an identity matrix. We define matrix $M$ as $\left(I'-Q\right)^{-1}$ and $A_{\mathcal{T}\times\left\{\mathcal{R}_{1} \cup \mathcal{R}_{2}\right\}} = {}MS$. For any transient state $\left(i, j\right)$ in $\mathcal{T}$ and recurrent states $\left(0, k\right)$ in $\mathcal{R}_{1}$ or $\left(k, 0\right)$ in $\mathcal{R}_{2}$, we define
\begin{center}
$\alpha_{\left(0, k\right)}\left(i, j\right) \coloneqq{}\mathbb{P}\{\,X_{n} = {}\left(0, k\right) \text{ for some } n \geq 0 \mid X_{0} = {} \left(i, j\right)\,\}$
$\alpha_{\left(k, 0\right)}\left(i, j\right) \coloneqq{}\mathbb{P}\{\,X_{n} = {}\left(k, 0\right) \text{ for some } n \geq 0 \mid X_{0} = {} \left(i, j\right)\,\}$
\end{center}
Then, the analysis in Lawler \cite{Lawler06} shows that $\alpha_{\left(0, k\right)}\left(i, j\right)$ is given by the entry corresponding to the transition from $\left(i, j\right)$ to $\left(0, k\right)$ in $A$ and $\alpha_{\left(k, 0\right)}\left(i, j\right)$ is given by the entry corresponding to the transition from $\left(i, j\right)$ to $\left(k, 0\right)$ in $A$. Then, for $k > 0\colon$
\begin{align}\label{eq:3.35}
p_{1}\left(u, w\right) & = {}1 - \sum_{k={}1}^{\infty}\alpha_{\left(k, 0\right)}\left(u, w\right)\nonumber\\
p_{2}\left(u, w\right) & = {}\sum_{k={}1}^{\infty}\alpha_{\left(k, 0\right)}\left(u, w\right)
\end{align}$\hfill \square$\\
\end{proof}

\textbf{Result \ref{Result3}}. The expected time $\phi_{\left(u, w\right)}$ is approximated by the solution to the following system of equations$\colon$
\begin{align}
    \begin{cases}
      \phi_{\left(u, w\right)} = {} 0 \hspace{6.93cm} \left(u, w\right) \in \mathcal{R}_{x}\nonumber\\
      \phi_{\left(u, w\right)} = {} 1 + \sum_{\left(i, j\right)} Q_{\left(u, w\right)\left(i, j\right)}\phi_{\left(i, j\right)} = {}0  \hspace{2.2cm} \left(u, w\right) \notin \mathcal{R}_{x}\\
    \end{cases}
\end{align}

\begin{proof}
 Let $\mathcal{S'} = {} \left\{\,\mathcal{R}_{x} \cup \mathcal{T} \,\right\}$ and $X'\left(t\right)$ be the modified CTMC of $X\left(t\right)$ $\left(\text{in Result 2}\right)$ with only states belonging to $\mathcal{S'}$ and let $q_{uw, ij}$ be the $\left(uw, ij\right)^{th}$ element of the generator $Q$ of the CTMC $ X'\left(t\right)$. As station $x$ becomes empty before $y$, there are no transitions from transient states $\left(i, j\right)$ to recurrent states in $\mathcal{R}_{y}$ which would make station $y$ empty $\Big(\left(0,k\right) \text{ for } y = {} 1 \text{ and } \left(k, 0\right) \text{ for } y = {} 2\Big)$. Given a set $\mathcal{R}_{x} \subseteq \mathcal{S'}$, the first-passage time $\left(\text{FPT}\right)$ to $\mathcal{R}_{x}$ is the random variable $T_{\mathcal{R}_{x}}$ defined by
\begin{center}
$T_{\mathcal{R}_{x}} = {} \inf\{\,t \geq 0 \mid X'\left(t\right) \in \mathcal{R}_{x}\,\}$
\end{center}
Let mean-first-passage time $\left(\text{MFPT}\right)$ to $\mathcal{R}_{x}$, starting at $\left(u, w\right) \in \mathcal{S'}$ be $\phi_{\left(u, w\right)}$, viz., $\phi_{\left(u, w\right)} = {} \mathbb{E}_{\left(u, w\right)}\left[T_{\mathcal{R}_{x}}\right]$. Clearly $\phi_{\left(u, w\right)} = {}0$ for $\left(u, w\right) \in \mathcal{R}_{x}$. Suppose $X'\left(0\right) = {}\left(u, w\right) \notin \mathcal{R}_{x}$. Let $J_{1}$ be the random variable denoting the time until the next transition and and let $Y_{1}$ be the state where it transitions to given $X'\left(0\right) = {}\left(u, w\right)$. We can calculate the MFPT by conditioning on the first transition, and then subtracting the time of the first transition. We have$\colon$
\begin{align} \label{eq:3.36}
\phi_{\left(u, w\right)} = & {} \displaystyle\mathop{\mathbb{E}}\Big[J_{1}\Big|X'\left(0\right) ={}\left(u, w\right)\Big] +\\\nonumber & \sum_{\left(i, j\right)}\displaystyle\mathop{\mathbb{E}}\Big[T_{\mathcal{R}_{x}} - J_{1}\Big|Y_{1} = {}\left(i, j\right), X'\left(0\right) = {}\left(u, w\right)\Big]\displaystyle\mathop{\mathbb{P}}\Big(Y_{1} = {}\left(i, j\right)| X'\left(0\right) = {}\left(u, w\right)\Big)
\end{align}
Using the Strong Markov property, we get that $\displaystyle\mathop{\mathbb{E}}\Big[T_{\mathcal{R}_{x}} - J_{1}\Big|Y_{1} = {}\left(i, j\right), X'\left(0\right) = {}\left(u, w\right)\Big] = {} \displaystyle\mathop{\mathbb{E}}\Big[T_{\mathcal{R}_{x}}|X'\left(0\right) = {}\left(i, j\right)\Big]$. We know that $\displaystyle\mathop{\mathbb{P}}\Big(Y_{1} = {}\left(i, j\right)| X'\left(0\right) = {}\left(u, w\right)\Big)$ is $\frac{q_{uw, ij}}{-q_{uw, uw}}$ and $\displaystyle\mathop{\mathbb{E}}\Big[J_{1}\Big|X'\left(0\right) ={}\left(u, w\right)\Big]$ is $-1/q_{uw, uw}$. By rearranging the expressions, we use the analysis in Durrett \cite{Durrett11} to get the following the equations$\colon$
\begin{align}
    \begin{cases}
      \phi_{\left(u, w\right)} = {} 0 \hspace{6.93cm} \left(u, w\right) \in \mathcal{R}_{x}\nonumber\\
      \phi_{\left(u, w\right)} = {} 1 + \sum_{\left(i, j\right)} Q_{\left(u, w\right)\left(i, j\right)}\phi_{\left(i, j\right)} = {}0  \hspace{2.2cm} \left(u, w\right) \notin \mathcal{R}_{x}\\
    \end{cases}
\end{align}
We estimate $ \phi_{\left(u, w\right)}$ by solving the above set of equations. $\hfill \square$\\
\end{proof}

\textbf{Result \ref{Result4}}. The probability $\displaystyle \mathop{\mathbb{P}}\Big(K_{i}={}k\Big)$, that exactly $k$ type $i$ customers are served at station 1 when $w$ type $i$ customers are completely served at station 2 is given by$\colon$
\begin{equation}\label{eq:3.37}
\displaystyle \mathop{\mathbb{P}}\Big(K_{i}={}k\Big)= {}\left(\frac{\mu_{i1}}{\mu_{i1} + \mu_{i2}}\right)^k \left(\frac{\mu_{i2}}{\mu_{i1} + \mu_{i2}}\right)^{w+k}\left[\binom{2k + w -1}{k} - \binom{2k + w -1}{k-1}\right]
\end{equation}

\begin{proof}
For a given $K = {}k$, this event requires that there were exactly $k$ service completions at station 1 and $w+k$ service completions at station 2, yielding a total of $2k+w$ service completions. The total number of ways by which this $k$ completion at station 1 can happen is $\binom{2k+w-1}{k}$. However, this include the events in which the queue length at station 2 reaches 0 before $k$ customers are served at station 1 as a result of which, the server at station 2 would switch before $w+k$
 customers are served at station 2. There are $\binom{2k+w-1}{k-1}$ such events where the queue length at station 2 reaches to 0 which should not be considered while determining the total number of events. Thus, the probability that event $K={}k$ occurs is $\left(\frac{\mu_{i1}}{\mu_{i1} + \mu_{i2}}\right)^k \left(\frac{\mu_{i2}}{\mu_{i1} + \mu_{i2}}\right)^{w+k}\left[\binom{2k + w -1}{k} - \binom{2k + w -1}{k-1}\right]$. $\hfill \square$\\
\end{proof}

\textbf{Result \ref{Result5}}. Let $W_{u, w}$ be the waiting time for the tagged customer queued at station 1 before it exits the system after getting served at station 2. Then$\colon$
\begin{align}
\displaystyle \mathop{\mathbb{E}}\left[W_{u, w}\right]={}
    \begin{cases}
       \mu_{1}^{-1} + \displaystyle\mathop{\mathbb{E}}\left[W_{u-1, 1}\right]\hspace{5.88cm} \text{ if } \left(u > 0, w ={}0\right)\\
       p\Big(\mu_{1}^{-1} + \left(w+1\right)\mu_{2}^{-1}\Big) + q\Big(\displaystyle\mathop{\mathbb{E}}\left[W_{0, w-1}\right]\Big)\hspace{2.0cm} \text{ if } \left(u = 0, w >0\right)\nonumber\\
       p\Big(\mu_{1}^{-1}+ \displaystyle\mathop{\mathbb{E}}\left[W_{u-1, w+1}\right]\Big) + q\Big(\displaystyle\mathop{\mathbb{E}}\left[W_{u, w-1}\right]\Big) \hspace{1.7cm} \text{ if } \left(u > 0, w > 0\right)\\
    \end{cases}
\end{align}
\vspace{3mm}
\begin{proof}
When \emph{u} is 0 and \emph{w} is 1, then the waiting time of the tagged customer is the sum of expected service time at station 1, expected waiting time at station 2, and expected service time at station 2, i.e., $\displaystyle\mathop{\mathbb{E}}\left[W_{0, 1}\right]={}\mu_{1}^{-1} + p\mu_{2}^{-1} + \mu_{2}^{-1}$ where $p = {}\frac{\mu_{1}}{\mu_{1} + \mu_{2}}$ as described in Result \ref{Result4}. The tagged customer waits at station 2 only when it gets served at station 1 before the customer at station 2, which happens with a probability \emph{p}. Similarly, when \emph{u} is 1 and \emph{w} is 0, then $\displaystyle\mathop{\mathbb{E}}\left[W_{1, 0}\right]={}2\mu_{1}^{-1} + p\mu_{2}^{-1} + \mu_{2}^{-1}$. Using the above results, we derive the waiting time relations for boundary conditions. When we have $\left(u > 0, w ={}0\right)$, the customer in service at station 1 waits for $\mu_{1}^{-1}$ duration before moving to station 2. Once the served customer moves to station 2, the additional waiting time for the tagged customer can be rewritten as $\displaystyle\mathop{\mathbb{E}}\left[W_{u-1, 1}\right]$. For the case when $\left(u = 0, w >0\right)$, with a probability $p$, the tagged customer gets served at station 1 before any customer gets served at station 2. The tagged customer moves to station 2 after waiting for $\mu_{1}^{-1}$ duration during its service at station 1 and waits for additional $\left(w+1\right)\mu_{2}^{-1}$ duration at station 2. With a probability $q$, there is a service completion at station 2 before the tagged customer gets served at station 1. The additional waiting time of the tagged customer can be rewritten as $ \displaystyle\mathop{\mathbb{E}}\left[W_{0, w-1}\right]$. Combining the above two relations for the boundary cases, we get the generalized recursive relation for $\displaystyle\mathop{\mathbb{E}}\left[W_{u, w}\right]$ when $\left(u > 0, w > 0\right)$. $\hfill \square$\\
\end{proof}

\textbf{Result \ref{Result6}}. Let $h_{1}\left(u\right)$ be the time to serve the $u$ customers at station 1 and $h_{2}\left(w\right)$ be the time to serve $w$ customers at station 2. Then $h_{1}\left(u\right)$ and $h_{2}\left(w\right)$ are random variables with gamma distribution with means $\frac{u}{\mu_{1}}$ and $\frac{w}{\mu_{2}}$ respectively. Subsequently for $\left(u,\text{ }w\right) \in \displaystyle\mathop{\mathbb{Z^{+}}}\colon$
\begin{equation}\label{eq:3.38}
\displaystyle \mathop{\mathbb{P}}\Big(h_{1}\left(u\right) < h_{2}\left(w\right)\Big) ={}\sum_{r={}u}^{\infty} \left(\frac{\mu_{1}}{\mu_{1} + \mu_{2}}\right)^r \left(\frac{\mu_{2}}{\mu_{1} + \mu_{2}}\right)^{w} \binom{r + w -1}{r}
\end{equation}
\begin{proof}
The probability, $\displaystyle\mathop {\mathbb{P}}\Big(h_{1}\left(u\right) < h_{2}\left(w\right)\Big)$, denotes the probability that the sum of $u$ exponential service times with parameter $\mu_{1}$ is less than the sum of $w$ exponential service times with parameter $\mu_{2}$. The probability that a service at station 1 completes before a service at station 2 is $p = {}\frac{\mu_{1}}{\mu_{1} + \mu_{2}}$. The service completions at the two stations form a sequence of $\operatorname{Bernoulli}$ random variables where $p$ denotes the probability that a customer at station 1 completes service prior to a customer at station 2 and $1-p$ denotes otherwise. Then, $\displaystyle \mathop{\mathbb{P}}\Big(h_{1}\left(u\right) < h_{2}\left(w\right)\Big)$ corresponds to where there are $u$ service completions at station 1 before the $w^{th}$ service completion at station 2,  i.e., if $Z \sim \operatorname{Negative} \operatorname{Binomial}(w; p)$, then $\displaystyle \mathop{\mathbb{P}}\Big(h_{1}\left(u\right) < h_{2}\left(w\right)\Big)$ is the probability that $Z \geq u$. $\hfill \square$\\
\end{proof}

\textbf{Result \ref{Result7}}. Given $g\left(u\right)$, $h\left(w\right)$, and $F_{h\left(w\right)}\left(t\right) = {}\int_{0}^{t}f_{h\left(w\right)}\left(x\right)dx$, the probability that station 2 empties before station 1 is given by Equation $\left(\ref{eq:3.39}\right)$.
\begin{equation}\label{eq:3.39}
\displaystyle \mathop{\mathbb{P}}\Big(h\left(w\right) < g\left(u\right)\Big) = {}\int_{0}^{\infty}\Big[\int_{0}^{t}f_{h\left(w\right)}\left(x\right)dx\Big]f_{g\left(u\right)}\left(t\right) dt
\end{equation}
\begin{proof}
The probability, $\displaystyle\mathop {\mathbb{P}}\Big(h\left(w\right) < g\left(u\right)\Big)$, denotes the probability that the sum of $w$ exponential service times with parameter $\mu_{2}$ is less than $u$ - busy periods of M/M/1 queue with Poisson arrivals with parameter $\lambda_{1}$ and exponential service times with parameter $\mu_{1}$.
\begin{align*} 
\displaystyle \mathop{\mathbb{P}}\Big(h\left(w\right) < g\left(u\right)\Big) & = {}\int_{-\infty}^{\infty}\displaystyle \mathop{\mathbb{P}}\Big\{h\left(w\right) < g\left(u\right)|g\left(u\right) = {}t\Big\}f_{g\left(u\right)}\left(t\right) dt\\
& = {}\int_{-\infty}^{\infty}\displaystyle \mathop{\mathbb{P}}\Big\{h\left(w\right) < t\Big\}f_{g\left(u\right)}\left(t\right) dt\\
& = {}\int_{0}^{\infty}F_{h\left(w\right)}\left(t\right)f_{g\left(u\right)}\left(t\right) dt\\
& = {}\int_{0}^{\infty}\Big[\int_{0}^{t}f_{h\left(w\right)}\left(x\right)dx\Big]f_{g\left(u\right)}\left(t\right) dt
\end{align*}$\hfill \square$\\\end{proof}
\section{Appendix B}\label{AppendixB}
\textbf{Waiting Time Analysis of Scenario$\colon m = {}2$}\\
In scenario $m = {}2$, when the tagged customer arrives at station 1 at time \emph{t} $=$ 0, it sees the server at station 1 serving queue 1, and the server at station 2 serving queue 2. The state of the system for scenario 2 at ${t = {}0}$ is represented as $\Big(L_{11}^{a}, L_{21}^{a}, 1, L_{12}^{a}, L_{22}^{a}, 2\Big)$. Figure \ref{fig:mesh3.15} shows the tree of events that can occur before this tagged customer departs from the system. 
\graphicspath {{Figures/}}
\begin{figure}[H]
\begin{center}
\includegraphics[scale=0.22]{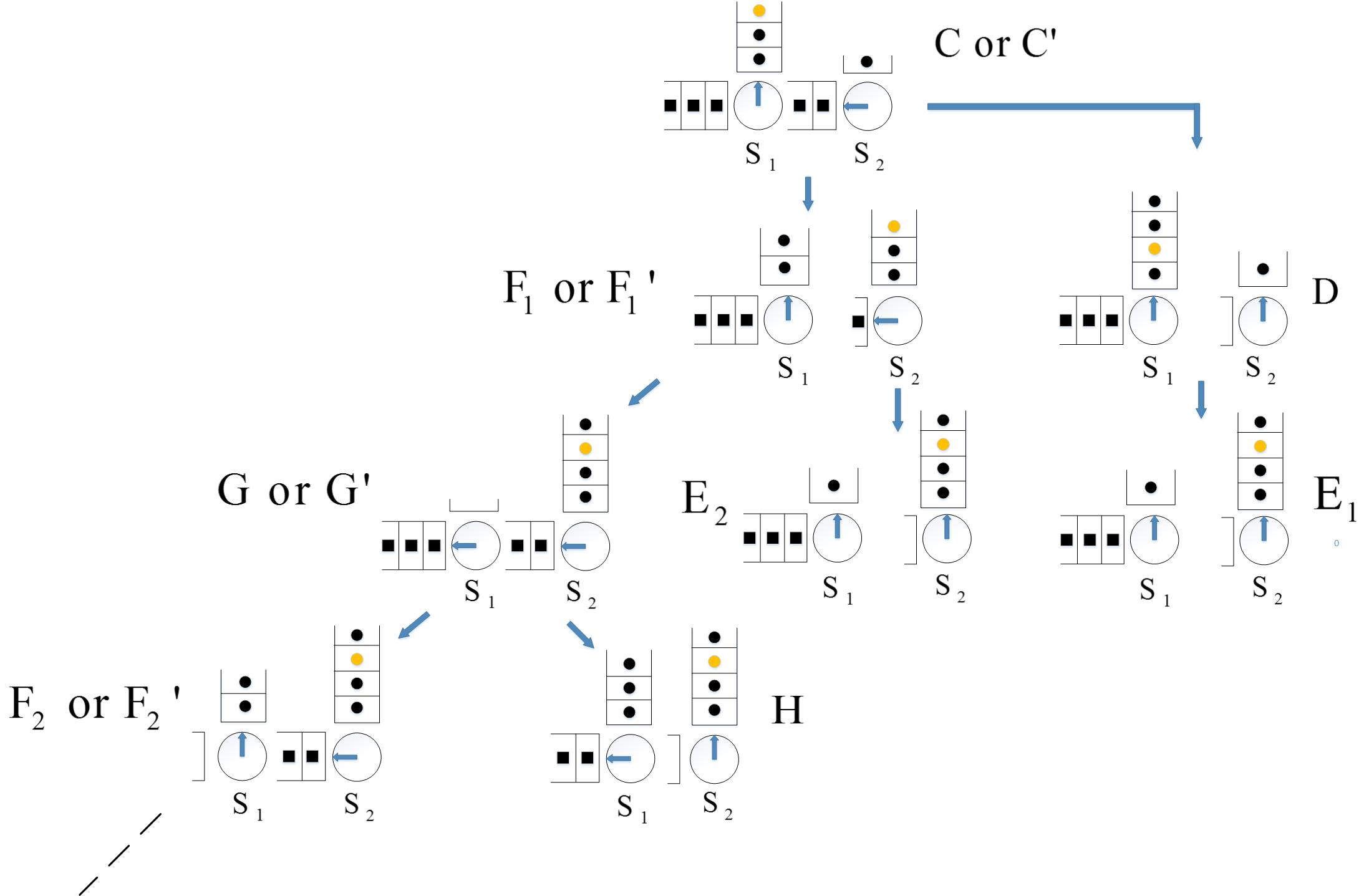}
\caption{Tree of different sub-scenarios for $m={}2$.}
\label{fig:mesh3.15}
\end{center}
\end{figure}
The corresponding sub-scenarios for scenario $m={}2$ are summarized in Table \ref{T:3.8}.\\
\renewcommand{\arraystretch}{1.2}
\begin{table}[H]
\centering
\caption{List of sub-scenarios of scenario $m={}2$.}\label{T:3.8}
\begin{tabular}{ |c|c|c| } 
\hline
Scenario & Sub-scenario & Event \\ 
\hline
$m={}2$ & $n = {}1   $ &  C $\prec$   D $\prec \text{E}_{1}$\\ 
$m={}2$ & $n = {}2   $ &  C' $\prec   \text{F}_{1}  \prec \text{E}_{2}$\\ 
$m={}2$ & $n = {}3   $ &  C' $\prec   \text{F'}_{1} \prec$ G $\prec$ H\\ 
$m={}2$ &  ...               &  ...\\ 
\hline
\end{tabular}
\end{table}
We observe that the system state on the arrival of tagged customer in scenario $m={}2$ is similar to system state at the beginning of event C or C', $t={}t_{C} \left(t_{C'}\right)$, in scenario $m={}1$ in terms of the queues being served at both the stations and the location of the tagged customer. Therefore, we can analyze scenario $m={}2$ using the same approach we used to analyze event C and the events that succeeded it in scenario $m={}1$. For instance, sub-scenario $\left(m={}2, n={}1\right)$  is identical to sub-scenario $\left(m={}1, n={}2\right)$ except that, at $t={}0$, the sub-scenario $\left(m={}2, n={}1\right)$ starts from event C and is followed by events D and $\text{E}_{1}$. Similarly, we analyze the other sub-scenarios for scenario $m={}2$ and determine the conditional waiting times and the probabilities of sub-scenarios to occur. The main difference in the analysis of scenario $ m = {}2$ from the analysis of scenario $m={}1$ is that $L_{12}^{'}\left(t_{E_{2}}\right)$ and $L_{12}^{'}\left(t_{H}\right)$ is $L_{11}^{a}+ L_{12}^{a}$ instead of $L_{11}^{a}- K_{1}$. We use this information when we determine the duration of the events $\text{E}_{2}$ and H in sub-scenarios $\left(m={}2, n={}2\right)$ and $\left(m={}2, n={}3\right)$. Once the conditional waiting times  and probabilities of different sub-scenarios have been determined, the waiting time $\displaystyle \mathop{\mathbb{E}}\left[W_{1}^{a}| m = {}2\right]$ can be calculated using Equation $\left(\ref{eq:3.40}\right)$.
\begin{equation}\label{eq:3.40}
\displaystyle \mathop{\mathbb{E}}\left[W_{1}^{a}| m = {}2\right]={}\sum_{s={}1}^{\infty}\displaystyle \mathop{\mathbb{E}}\left[W_{1}^{a}| m = {}2, n={}s\right]\Pr\left(n={}s| m={}2\right)
\end{equation}
\section{Appendix C}\label{AppendixC}
\textbf{Waiting Time Analysis of Scenario$\colon m = {}3$}\\
In scenario $m = {}3$, when the tagged customer arrives at station 1 at time \emph{t} $=$ 0, it sees the server at station 1 serving queue 2 and the server at station 2 serving queue 1. The state of the system for scenario 3 at ${t = {}0}$ is represented as $\Big(L_{11}^{a}, L_{21}^{a}, 2, L_{12}^{a}, L_{22}^{a}, 1\Big)$. Figure \ref{fig:mesh3.16} shows the tree of events that can occur before the tagged customer departs from the system.\\
\graphicspath {{Figures/}}
\begin{figure}[H]
\begin{center}
\includegraphics[scale=0.13]{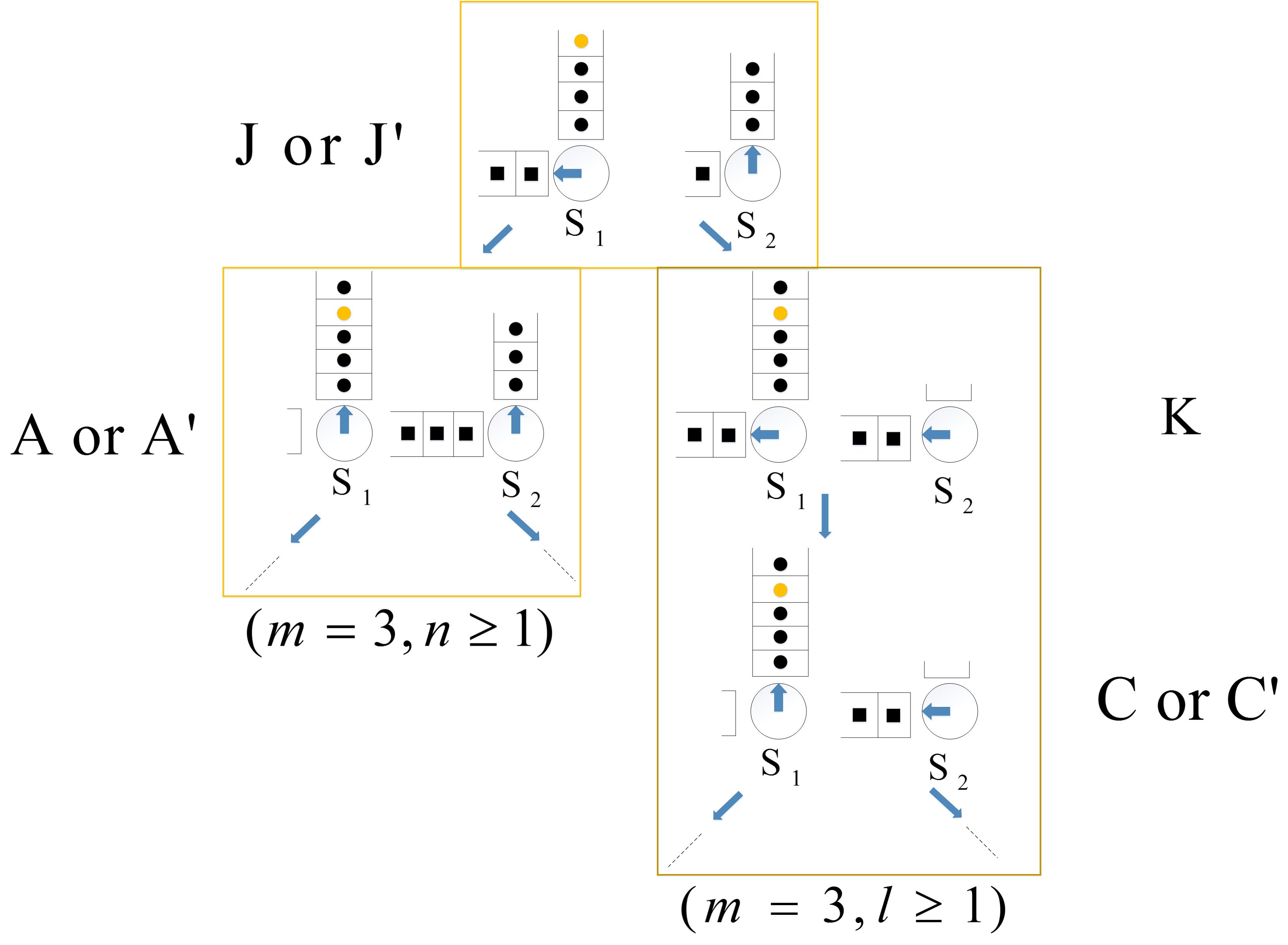}
\caption{Tree of different sub-scenarios for $m={}3$.}
\label{fig:mesh3.16}
\end{center}
\end{figure}
The corresponding sub-scenarios for scenario $m={}3$ are summarized in Table \ref{T:3.9}.\\
\renewcommand{\arraystretch}{1.2}
\begin{table}[h!]
\centering
\caption{List of sub-scenarios of scenario $m={}3$.}\label{T:3.9}
\begin{tabular}{ |c|c|c| } 
\hline
Scenario & Sub-scenario & Event \\ 
\hline
$m={}3$ & $n = {}1   $ &  J $\prec$ A  $\prec$ B\\ 
$m={}3$ & $n = {}2   $ &  J $\prec$ A' $\prec$ C $\prec$   D $\prec \text{E}_{1}$\\ 
$m={}3$ & $n = {}3   $ &  J $\prec$ A' $\prec$ C' $\prec   \text{F}_{1}  \prec \text{E}_{2}$\\ 
$m={}3$ & $n = {}4   $ &  J $\prec$ A' $\prec$ C' $\prec   \text{F'}_{1} \prec$ G $\prec$ H\\ 
$m={}3$ &  ...               &  ...\\ 
$m={}3$ & $l = {}1    $ &  J' $\prec$ K $\prec$ C  $\prec$   D $\prec \text{E}_{1}$\\ 
$m={}3$ & $l = {}2    $ &  J' $\prec$ K $\prec$ C' $\prec   \text{F}_{1}  \prec \text{E}_{2}$\\ 
$m={}3$ & $l = {}3    $ &  J' $\prec$ K $\prec$ C' $\prec   \text{F'}_{1} \prec$ G $\prec$ H\\ 
$m={}3$ &  ...               &  ...\\ 
\hline
\end{tabular}
\end{table}

We derive the mean waiting times using a similar analysis of the sub-scenarios. Once the conditional waiting times  and probabilities of different sub-scenarios have been determined, the waiting time $\displaystyle \mathop{\mathbb{E}}\left[W_{1}^{a}| m = {}3\right]$ can be calculated using Equation $\left(\ref{eq:3.41}\right)$.
\begin{equation}\label{eq:3.41}
\begin{aligned}
\displaystyle \mathop{\mathbb{E}}\left[W_{1}^{a}| m = {}3\right]= &{}\sum_{r={}1}^{\infty}\displaystyle \mathop{\mathbb{E}}\left[W_{1}^{a}| m = {}3, n={}r\right]\Pr\left(n={}r| m={}3\right) + \\
& \sum_{s={}1}^{\infty}\displaystyle \mathop{\mathbb{E}}\left[W_{1}^{a}| m = {}3, l={}s\right]\Pr\left(l={}s| m={}3\right)
\end{aligned}
\end{equation}
We next determine the conditional waiting times for sub-scenarios $\left(m = {}3, n \geq 1\right)$ and $\left(m = {}3, l \geq 1\right)$.\\

\textbf{Waiting Time Analysis of Sub-scenarios$\colon\left(m = {}3, n \geq 1\right)$}\\
As we analyze sub-scenarios for $\left(m = {}3, n \geq 1\right)$, we observe that the systems state after event J is similar to systems state at the beginning of scenario $m={}1$, i.e., at $t={}t_{A} \text{ or } \left(t_{A'}\right)$ $\left(\text{see Figure } \ref{fig:mesh3.17}\right)$. We can analyze sub-scenarios $\left(m = {}3, n \geq 1\right)$ by analyzing event J and then using the same approach used to analyze sub-scenarios for $m = {}1$.  Next, we derive the expressions for the expected waiting times for event J.\\

In event J, the $L_{21}^{a}$ customers of queue 2 at station 1 and additional external arrivals are served before all $L_{12}^{a}$ customers of queue 1 at station 2. Let $g_{21}\left(L_{21}^{a}\right)$ denote the hitting time to 0, $T_{J}$, for queue 2 at station 1 with $L_{21}^{a}$ customers. The probability, $\displaystyle \mathop{\mathbb{P}}\Big(\text {Event $\text{J}$}\Big)$, that $L_{21}^{a}$ customers of queue 2 at station 1 and additional external arrivals are served before $L_{12}^{a}$ customers of queue 1 at station 2 is determined using Result $\ref{Result7}$ and is given by Equation $\left(\ref{eq:3.42}\right)$.
\begin{equation}\label{eq:3.42}
\displaystyle\mathop{\mathbb{P}}\Big(\text{Event J}\Big) ={}\displaystyle \mathop{\mathbb{P}}\Big(g_{21}\left(L_{21}^{a}\right)<{} h_{12}\left(L_{12}^{a}\right)\Big)
\end{equation}

\graphicspath {{Figures/}}
\begin{figure}[H]
\begin{center}
\includegraphics[scale=0.18]{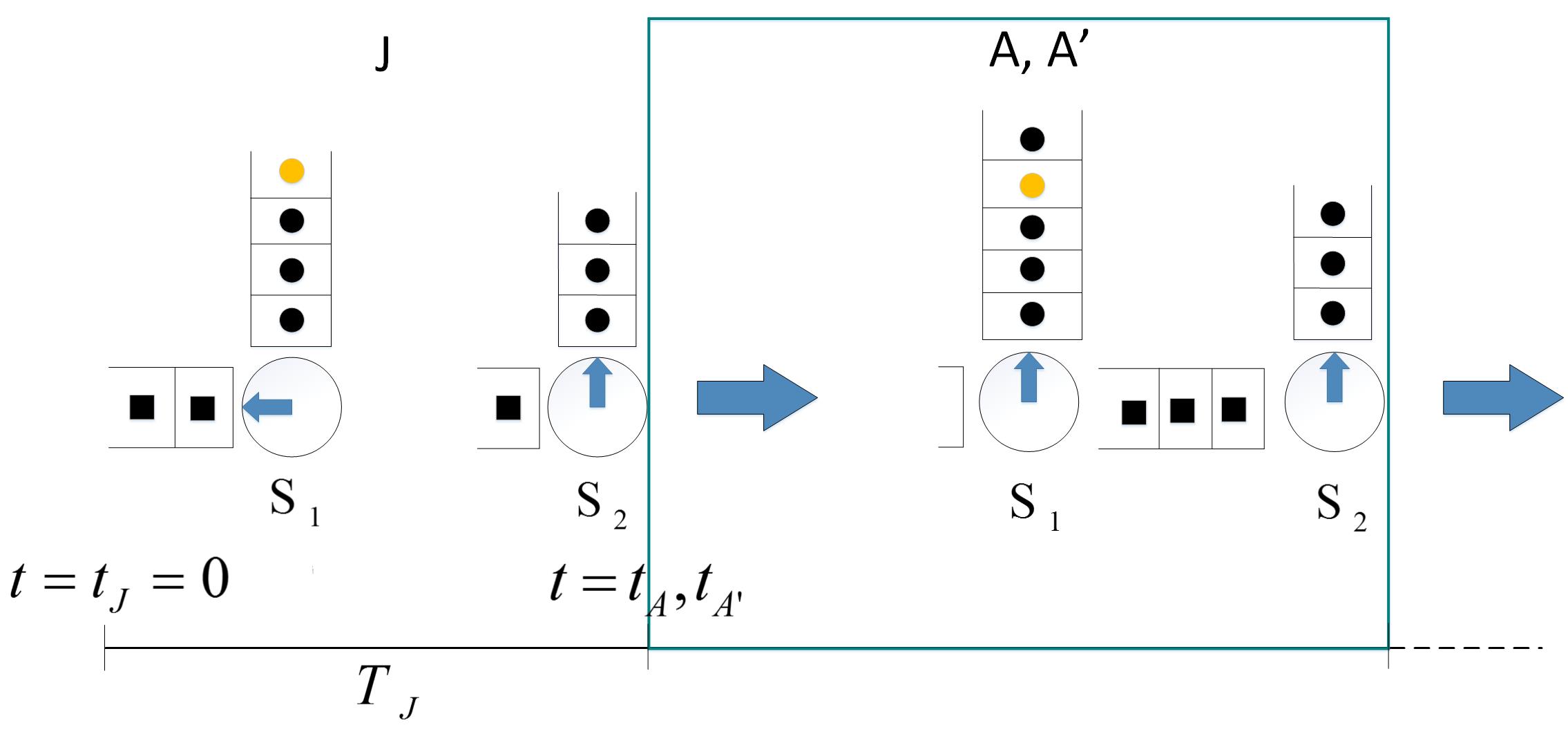}
\caption{Sequence of events for sub-scenario with $m = {}3$, $n \geq 1$.}
\label{fig:mesh3.17}
\end{center}
\end{figure}

The queue lengths at $t={}t_{A}$ or $t={}t_{A'}$ are given by  Equation $\left(\ref{eq:3.43}\right)$ where the upper bound on $V_{5}$ is $L_{12}^{a}-1$.
\begin{equation}\label{eq:3.43}
\begin{aligned}
L_{11}\left(t_{A}\right)=  & {} L_{11}^{a} + 1 + S_{1}\left(T_{J}\right)\\
L_{12}\left(t_{A}\right)=  & {} L_{12}^{a} - V_{5}\\
L_{21}\left(t_{A}\right)=  & {} 0\\
L_{22}\left(t_{A}\right)=  & {}L_{22}^{a}+L_{21}^{a}+S_{2}\left(T_{J}\right)\\
L_{11}'\left(t_{A}\right)= & {} L_{11}^{a}
\end{aligned}
\end{equation}

Now, we can analyze remaining events of sub-scenarios $\left(m = {}3, n \geq 1\right)$ using the same approach we used to analyze sub-scenarios for scenario $m={}1$. We analyze all the sub-scenarios for scenario $\left(m = {}3, n \geq 1\right)$ and determine the conditional waiting times $\displaystyle\mathop{\mathbb{E}}\left[W_{1}^{a}| m = {}3, n \geq{}1\right]$  and the probabilities of sub-scenarios $\Pr\left(n\geq{}1| m={}3\right)$ using Equation $\left(\ref{eq:3.34}\right)$.\\

\textbf{Waiting Time Analysis of Sub-scenarios$\colon\left(m = {}3, l \geq 1\right)$}\\
On analyzing sub-scenarios for $\left(m = {}3, l \geq 1\right)$, we observe that possible sequence of events after events J' and K is similar to possible sequence of events at the beginning of scenario $m={}2$, i.e., at $t={}t_{C} \text{ or } \left(t_{C'}\right)$ $\left(\text{see Figure } \ref{fig:mesh3.18}\right)$. Therefore, we can analyze sub-scenarios $\left(m = {}3, l \geq 1\right)$ by analyzing events J' and K and then using the same approach used to analyze sub-scenarios for scenario $m = {}2$.  Next, we derive the expressions for the expected waiting times for events J' and K.\\

In event J', all $L_{12}^{a}$ customers of queue 1 at station 2 are served before the $L_{21}^{a}$ customers of queue 2 at station 1 and additional external arrivals. The time, $T_{J'}$, it takes to serve all customers of queue 1 at station 2 is $h_{12}\left(L_{12}^{a}\right)$. The probability, $\displaystyle \mathop{\mathbb{P}}\Big(\text {Event $\text{J'}$}\Big)$, that all $L_{12}^{a}$ customers of queue 1 at station 2 are served before the $L_{21}^{a}$ customers of queue 2 at station 1 and additional external arrivals is given by Equation $\left(\ref{eq:3.44}\right)$.
\begin{equation}\label{eq:3.44}
\displaystyle\mathop{\mathbb{P}}\Big(\text{Event J'}\Big) ={}\displaystyle \mathop{\mathbb{P}}\Big({}h_{12}\left(L_{12}^{a}\right) < g_{21}\left(L_{21}^{a}\right)\Big)
\end{equation}

\graphicspath {{Figures/}}
\begin{figure}[h!]
\begin{center}
\includegraphics[scale=0.15]{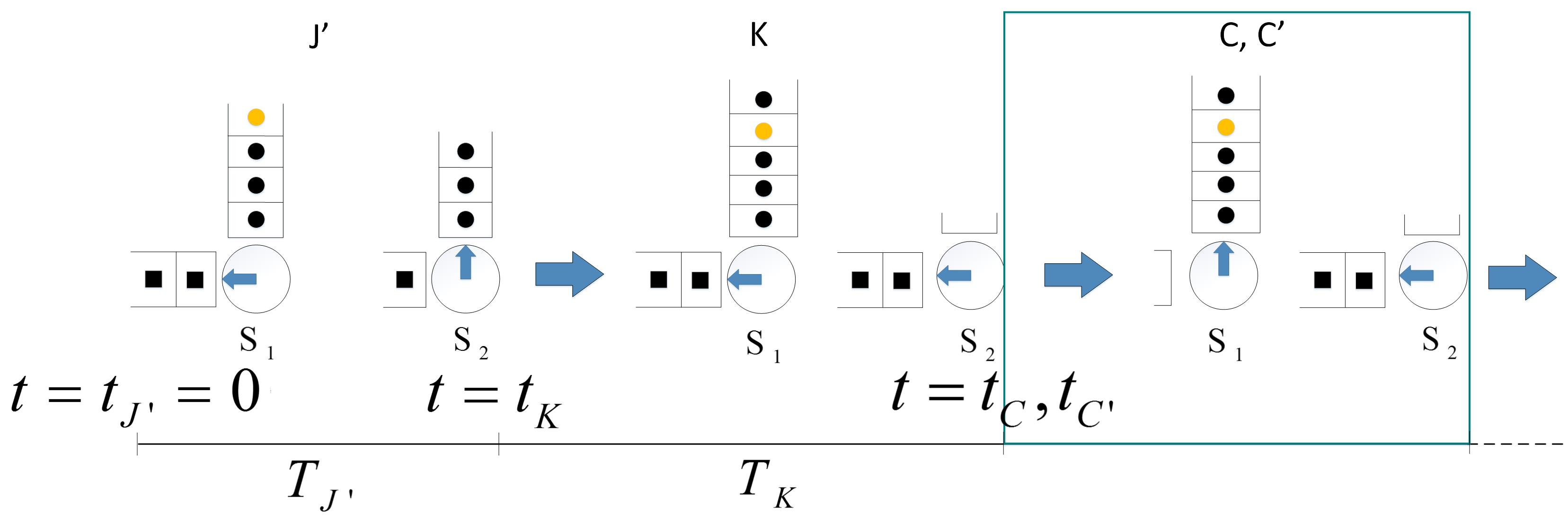}
\caption{Sequence of events for sub-scenario with $m = {}3$, $l \geq 1$.}
\label{fig:mesh3.18}
\end{center}
\end{figure}

In event K, the $L_{21}\left(t_{K}\right)$ customers of queue 2 at station 1 and additional external arrivals are served before $L_{22}\left(t_{K}\right)$  customers of queue 2 at station 2 and additional arrivals from station 1. The probability, $\displaystyle\mathop{\mathbb{P}}\Big(\text{Event K}\Big)$, is 1 since queue 1 at station 2 is empty during event K due to which, the server at station 2 will not switch to queue 1 once queue 2 is empty. During the time $\left(0, t_{J'}\right)$, additional $S_{2}\left(T_{J'}\right)$ type 2 customers arrive at station 1 and $V_{6}$ type 2 customers leave station 1. The upper bound on $V_{6}$ is $L_{21}^{a}+S_{2}\left(T_{J'}\right)-1$. Thus, $L_{21}\left(t_{K}\right)= {} L_{21}^{a}+ S_{2}\left(T_{J'}\right)-V_{6}$ . The hitting time to 0, $T_{K}$, for queue 2 at station 1 with $L_{21}\left(t_{K}\right)$ customers is $g_{21}\Big(L_{21}\left(t_{K}\right)\Big)$.\\

The queue lengths at $t={}t_{C}$ or $t={}t_{C'}$ are given by Equation $\left(\ref{eq:3.45}\right)$ where the upper bound on $V_{7}$ is $L_{22}^{a}+L_{21}^{a}+S_{2}\left(T_{J'}+T_{K}\right)-1$.
\begin{equation}\label{eq:3.45}
\begin{aligned}
L_{11}\left(t_{C}\right) & =  {} L_{11}^{a}+1+S_{1}\left(T_{J'}+T_{K}\right)\\
L_{12}\left(t_{C}\right) & =  {} 0\\
L_{21}\left(t_{C}\right) & =  {} 0\\
L_{22}\left(t_{C}\right) & =  {}L_{22}^{a}+L_{21}^{a}+S_{2}\left(T_{J'}+T_{K}\right)-V_{7}\\
L_{11}'\left(t_{C}\right) & = {} L_{11}^{a}\\
\end{aligned}
\end{equation}
Now, we analyze remaining events of sub-scenarios $\left(m = {}3, l \geq 1\right)$ using the same approach we used to analyze sub-scenarios for scenario $m={}2$. We analyze all the sub-scenarios for scenario $\left(m = {}3, l \geq 1\right)$ and determine the conditional waiting times $\displaystyle\mathop{\mathbb{E}}\left[W_{1}^{a}| m = {}3, l \geq{}1\right]$  and the probabilities of sub-scenarios $\Pr\left(l\geq{}1| m={}3\right)$ using Equation $\left(\ref{eq:3.34}\right)$.
\section{Appendix D}\label{AppendixD}
\textbf{Waiting Time Analysis of Scenario$\colon m = {}4$}\\
In scenario $m = {}4$, when the tagged customer arrives at station 1 at time \emph{t} $=$ 0, it sees the server at both the stations serving queue 2. The state of the system for scenario 4 at ${t = {}0}$ is $\Big(L_{11}^{a}, L_{21}^{a}, 2, L_{12}^{a}, L_{22}^{a}, 2\Big)$. Figure \ref{fig:mesh3.19} shows the tree of events that can occur before the tagged customer departs from the system.
\graphicspath {{Figures/}}
\begin{figure}[H]
\begin{center}
\includegraphics[scale=0.15]{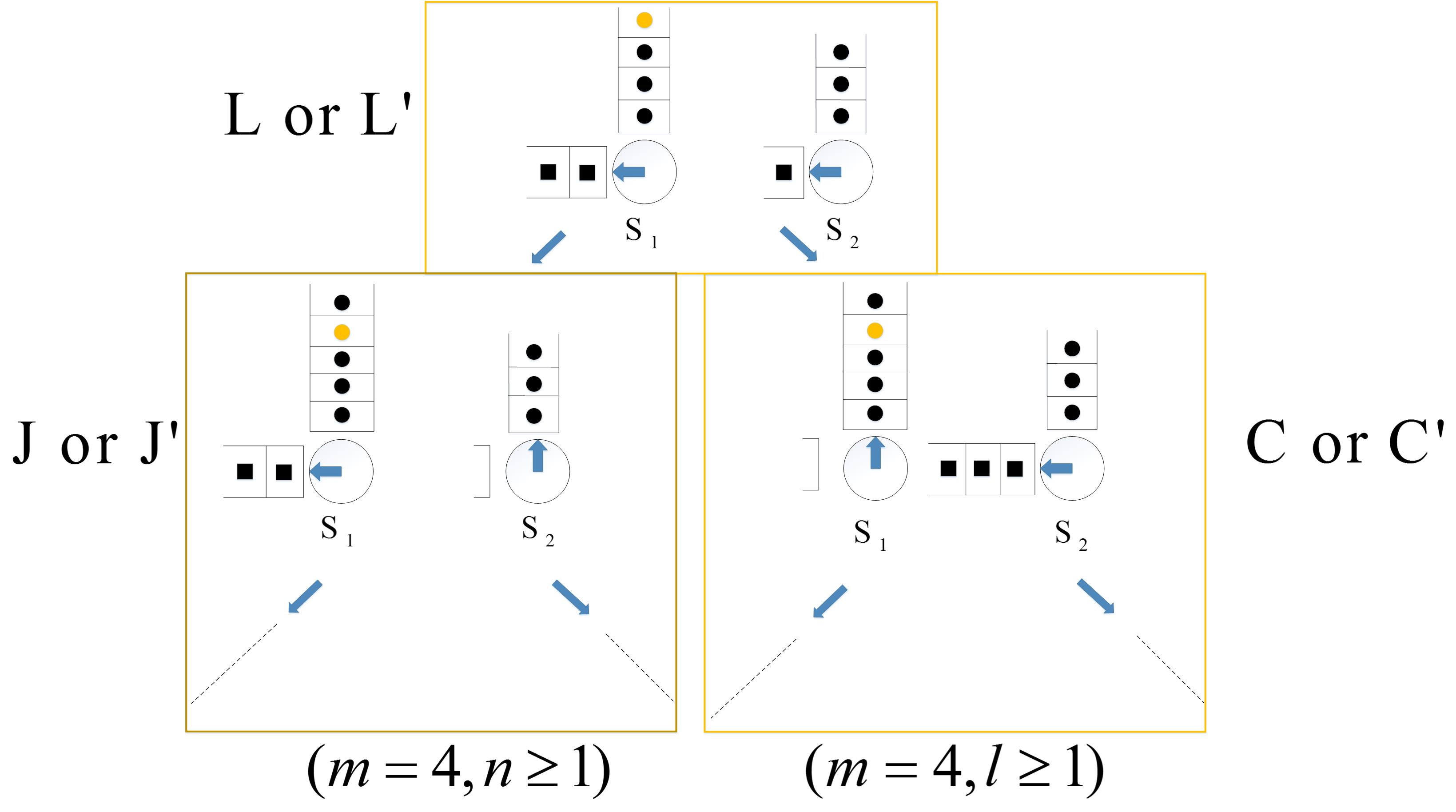}
\caption{Tree of different sub-scenarios for $m={}4$.}
\label{fig:mesh3.19}
\end{center}
\end{figure}
The corresponding sub-scenarios of scenario $m={}4$ are summarized in Table \ref{T:3.10}.\\
\begin{table}[h!]
\centering
\caption{List of sub-scenarios of scenario $m={}4$.}\label{T:3.10}
\begin{tabular}{ |c|c|c| } 
\hline
Scenario & Sub-scenario & Event \\ 
\hline
$m={}4$ & $n = {}1   $ &  L $\prec$ J $\prec$ A  $\prec$ B\\ 
$m={}4$ & $n = {}2   $ &  L $\prec$ J $\prec$ A' $\prec$ C $\prec$   D $\prec \text{E}_{1}$\\ 
$m={}4$ & $n = {}3   $ &  L $\prec$ J $\prec$ A' $\prec$ C' $\prec   \text{F}_{1}  \prec \text{E}_{2}$\\ 
$m={}4$ & $n = {}4   $ &  L $\prec$ J $\prec$ A' $\prec$ C' $\prec   \text{F'}_{1} \prec$ G $\prec$ H\\ 
$m={}4$ & $  \vdots   $ &    $\vdots$ \\ 
$m={}4$ & $l = {}1    $ &  L' $\prec$ C  $\prec$   D $\prec \text{E}_{1}$\\ 
$m={}4$ & $l = {}2    $ &  L' $\prec$ C' $\prec   \text{F}_{1}  \prec \text{E}_{2}$\\ 
$m={}4$ & $l = {}3    $ &  L' $\prec$ C' $\prec   \text{F'}_{1} \prec$ G $\prec$ H\\ 
$m={}4$ & $  \vdots   $ &     $\vdots$ \\ 
\hline
\end{tabular}
\end{table}

We derive the mean waiting times using a similar analysis of the sub-scenarios. Once the conditional waiting times  and probabilities of different sub-scenarios have been determined, the waiting time $\displaystyle \mathop{\mathbb{E}}\left[W_{1}^{a}| m = {}4\right]$ can be calculated using Equation $\left(\ref{eq:3.46}\right)$.
\begin{equation}\label{eq:3.46}
\begin{aligned}
\displaystyle \mathop{\mathbb{E}}\left[W_{1}^{a}| m = {}4\right]= & {}\sum_{r={}1}^{\infty}\displaystyle \mathop{\mathbb{E}}\left[W_{1}^{a}| m = {}4, n={}r\right]\Pr\left(n={}r| m={}4\right) + \\
& \sum_{s={}1}^{\infty}\displaystyle \mathop{\mathbb{E}}\left[W_{1}^{a}| m = {}4, l={}s\right]\Pr\left(l={}s| m={}4\right)
\end{aligned}
\end{equation}
We next determine the conditional waiting times for sub-scenarios $\left(m = {}4, n \geq 1\right)$ and $\left(m = {}4, l \geq 1\right)$.\\

\textbf{Waiting Time Analysis of Sub-scenario$\colon\left(m = {}4, n \geq 1\right)$}\\
As we analyze sub-scenarios for $\left(m = {}4, n \geq 1\right)$, we observe that the sequence of events after event L is similar to sequence of events at the beginning of scenario $m={}3$, i.e., at $t={}t_{J} \text{ or } \left(t_{J'}\right)$ $\left(\text{see Figure } \ref{fig:mesh3.20}\right)$. We can analyze sub-scenarios $\left(m = {}4, n \geq 1\right)$ by analyzing event L and then using the same approach used to analyze sub-scenarios for scenario $m = {}3$.\\

Next, we derive the expressions for the expected waiting times for event L. In event L, the $L_{22}^{a}$ customers of queue 2 at station 2 and additional arrivals from station 1 are served before the $L_{21}^{a}$ existing customers of queue 2 at station 1 and additional external arrivals. The hitting time to 0,  $T_{L}$, for queue 2 with $L_{22}^{a}$ customers at station 2 and $L_{21}^{a}$ type 2 customers at station 1 is $\phi_{\left(L_{21}^{a}, L_{22}^{a}\right)}$ and is determined using Result $\ref{Result3}$ where $x={}2$ and $y={}1$. The probability, $\displaystyle \mathop{\mathbb{P}}\Big(\text {Event $\text{L}$}\Big)$, that  the $L_{22}^{a}$ customers of queue 2 at station 2 and additional arrivals from station 1 are served before the $L_{21}^{a}$ existing customers of queue 2 at station 1 and additional external arrivals is given by Equation $\left(\ref{eq:3.47}\right)$ and is determined using Result $\ref{Result2}$.

\begin{equation}\label{eq:3.47}
\displaystyle\mathop{\mathbb{P}}\Big(\text{Event $\text{L}$}\Big) ={}\sum_{k={}1}^{\infty}\alpha_{\left(k, 0\right)}\left(L_{21}^{a}, L_{22}^{a}\right)
\end{equation}

\graphicspath {{Figures/}}
\begin{figure}[H]
\begin{center}
\includegraphics[scale=0.15]{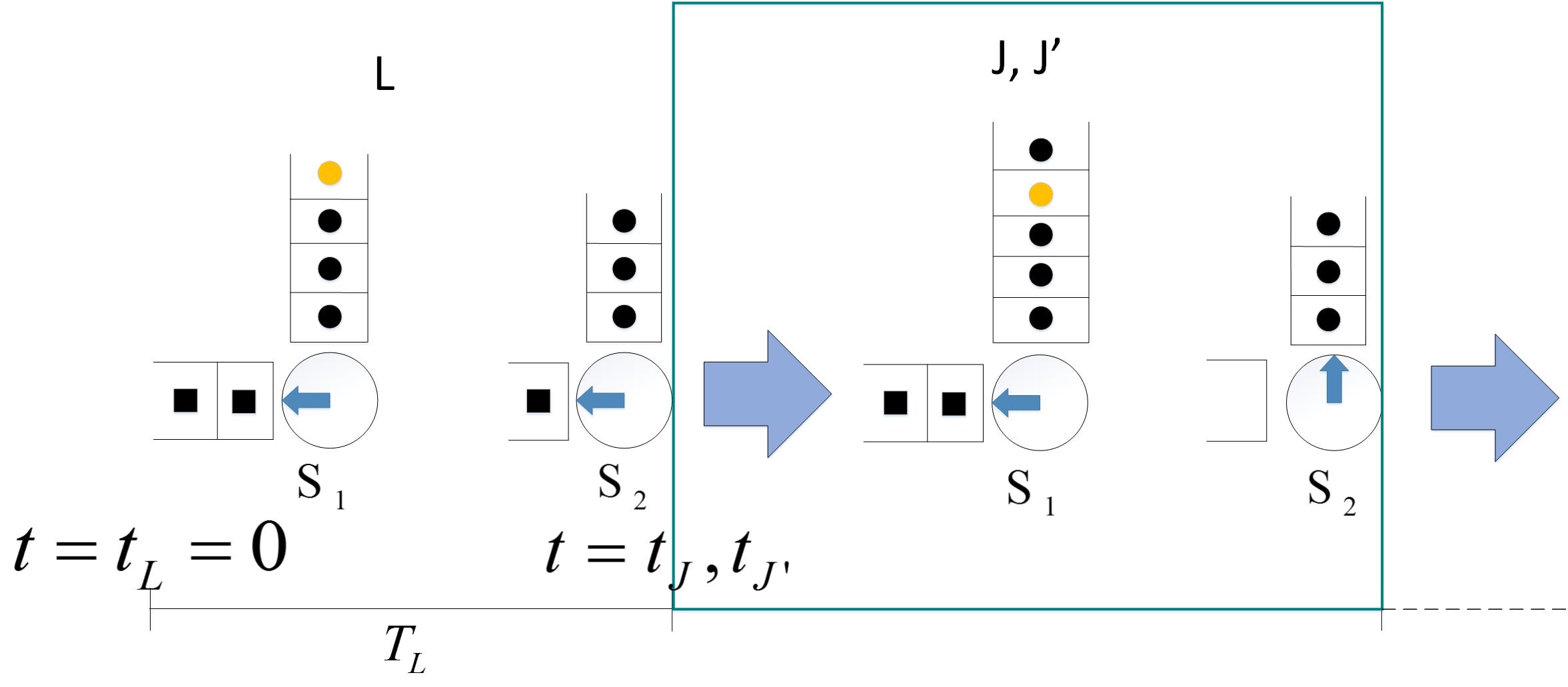}
\caption{Sequence of events for sub-scenario with $m = {}4$, $n \geq 1$.}
\label{fig:mesh3.20}
\end{center}
\end{figure}

The queue lengths at $t={}t_{J}$ or $t={}t_{J'}$  are given by Equation $\left(\ref{eq:3.41}\right)$  where the upper bound on $V_{8}$ is $L_{21}^{a}+S_{2}\left(T_{L}\right)-1$.
\begin{equation}\label{eq:3.48}
\begin{aligned}
L_{11}\left(t_{L}\right) & = {} L_{11}^{a}+1+S_{1}\left(T_{L}\right)\\
L_{12}\left(t_{L}\right) & = {} L_{12}^{a}\\
L_{21}\left(t_{L}\right) & = {}L_{21}^{a}+S_{2}\left(T_{L}\right)-V_{8}\\
L_{22}\left(t_{L}\right) & = {}0\\
L_{11}'\left(t_{L}\right) & ={} L_{11}^{a}\\
\end{aligned}
\end{equation}
Now, we analyze remaining events of sub-scenarios $\left(m = {}4, n \geq 1\right)$ using the same approach we used to analyze sub-scenarios for scenario $m={}3$. We analyze all the sub-scenarios for scenario $\left(m = {}4, n \geq 1\right)$ and determine the conditional waiting times $\displaystyle\mathop{\mathbb{E}}\left[W_{1}^{a}| m = {}4, n \geq{}1\right]$  and the probabilities of sub-scenarios $\Pr\left(n\geq{}1| m={}4\right)$ using Equation $\left(\ref{eq:3.46}\right)$.\\

\textbf{Waiting Time Analysis of Sub-scenario$\colon\left(m = {}4, l \geq 1\right)$}\\
On analyzing sub-scenarios for $\left(m = {}4, l \geq 1\right)$, we observe that the systems state after event L' is similar to systems state at the beginning of scenario $m={}2$, i.e., at $t={}t_{C} \text{ or } \left(t_{C'}\right)$ $\left(\text{see Figure } \ref{fig:mesh3.21}\right)$. We can analyze sub-scenarios $\left(m = {}4, l \geq 1\right)$ by analyzing event L' and then using the same approach used to analyze sub-scenarios for scenario $m = {}2$.  Next, we derive the expressions for the expected waiting times for events L'.\\

In event L', the $L_{21}^{a}$ customers of queue 2 at station 1 and additional external arrivals are served before all $L_{22}^{a}$ customers of queue 2 at station 2 and additional arrivals from station 1. The hitting time to 0, $T_{L'}$, for queue 2 at station 1 with $L_{21}^{a}$ customers is $g_{21}\left(L_{21}^{a}\right)$. The probability, $\displaystyle \mathop{\mathbb{P}}\Big(\text {Event $\text{L'}$}\Big)$, that the $L_{21}^{a}$ customers of queue 2 at station 1 and additional external arrivals are served before all $L_{22}^{a}$ customers of queue 2 at station 2 and additional arrivals from station 1 is given by Equation $\left(\ref{eq:3.49}\right)$ and is determined using Result $\ref{Result2}$.
\begin{equation}\label{eq:3.49}
\displaystyle\mathop{\mathbb{P}}\Big(\text{Event $\text{L'}$}\Big) ={}\sum_{k={}1}^{\infty}\alpha_{\left(0, k\right)}\left(L_{21}^{a}, L_{22}^{a}\right)
\end{equation}

\graphicspath {{Figures/}}
\begin{figure}[H]
\begin{center}
\includegraphics[scale=0.18]{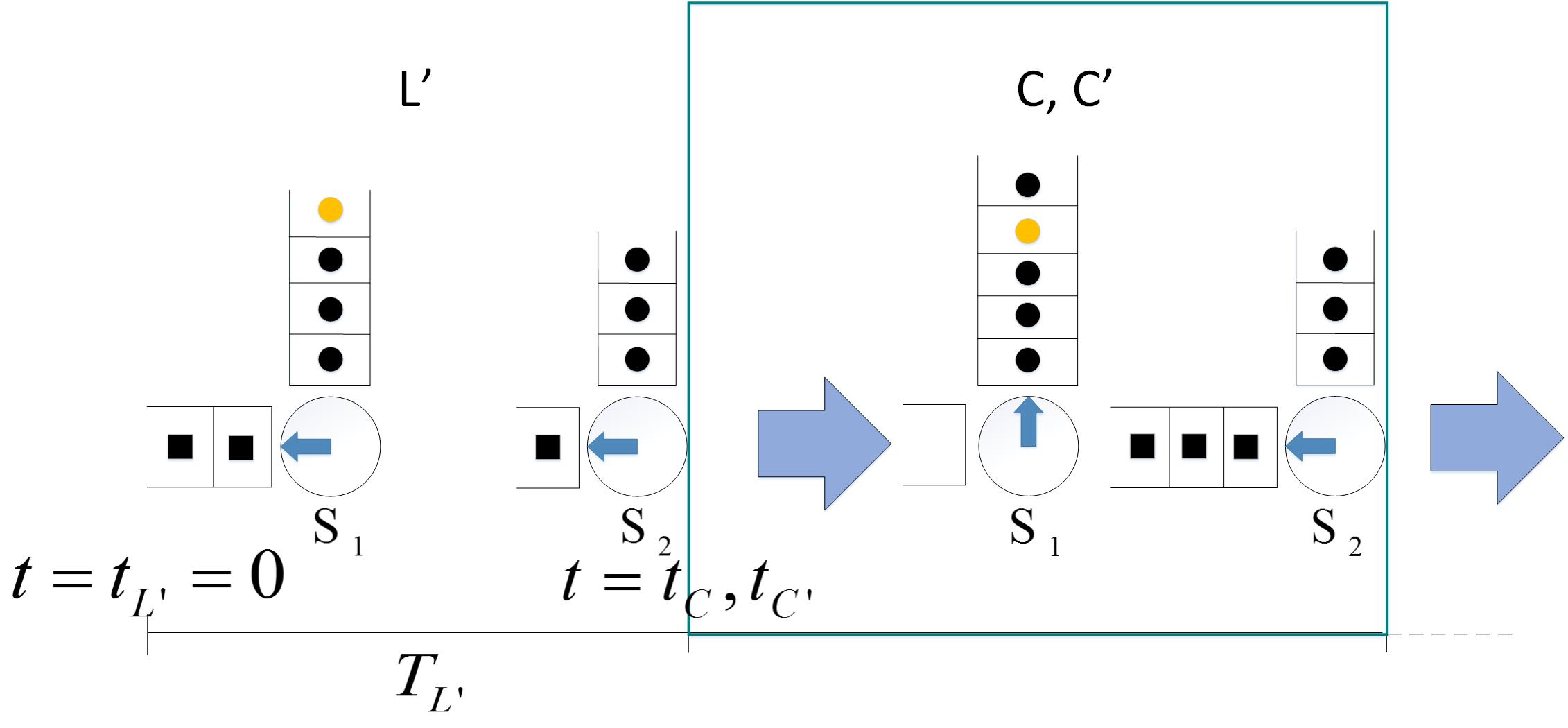}
\caption{Sequence of events for sub-scenario with $m = {}3$, $l \geq 1$.}
\label{fig:mesh3.21}
\end{center}
\end{figure}

The queue lengths at $t={}t_{C}$ or $t={}t_{C'}$ are given by Equation $\left(\ref{eq:3.50}\right)$  where the upper bound on $V_{9}$ is $L_{22}^{a}+L_{21}^{a}+S_{2}\left(T_{L'}\right)-1$.\\
\begin{equation}\label{eq:3.50}
\begin{aligned}
L_{11}\left(t_{C}\right)=  & {} L_{11}^{a}+1+S_{1}\left(T_{L'}\right)\\
L_{12}\left(t_{C}\right)=  & {} L_{12}^{a}\\
L_{21}\left(t_{C}\right)=  & {} 0\\
L_{22}\left(t_{C}\right)=  & {}L_{22}^{a}+L_{21}^{a}+S_{2}\left(T_{L'}\right)-V_{9}\\
L_{11}'\left(t_{C}\right)= & {} L_{11}^{a}
\end{aligned}
\end{equation}
Now, we can analyze remaining events of sub-scenarios $\left(m = {}4, l \geq 1\right)$ using the same approach we used to analyze scenario $m={}2$. We analyze all the sub-scenarios for scenario $\left(m = {}4, l \geq 1\right)$ and determine the conditional waiting times $\displaystyle\mathop{\mathbb{E}}\left[W_{1}^{a}| m = {}4, l \geq{}1\right]$  and the probabilities of sub-scenarios $\Pr\left(l\geq{}1| m={}4\right)$.

\begin{thebibliography}{19}
\bibitem{Bertsimas99}
Bertsimas, D., \& Mourtzinou, G. (1999). \newblock ``Decomposition Results for General Polling Systems and
their Applications'' \newblock \emph{Queueing Systems}, vol 31: 3, pp. 295 -- 316.

\bibitem{Boona11}
Boona, M.M.A., van der Mei, R.D., \&  Winands, E.M.M. (2011). \newblock ``Applications of Polling Systems'' \newblock \emph{Surveys in Operations Research and Management Science}, vol 16: 2, pp. 67 -- 82.

\bibitem{Borst97}
Borst, S.C., \& Boxma, O. (1997). \newblock ``Polling Models With and Without Switchover Times'', \newblock \emph{Operations Research}, vol 45: 4, pp. 536 -- 543.

\bibitem{Boxma87}
Boxma, O.J., \& Groenendijk, W. P. (1987). \newblock ``Pseudo-Conservation Laws in Cyclic-Service Systems'' \newblock \emph{Journal of Applied Probability}, vol 24: 4, pp. 949 -- 964.

\bibitem{Boxma09}
Boxma, O., Bruin, J., \& Fralix, B. (2009). \newblock ``Sojourn Times in Polling Systems With Various Service Disciplines'', \newblock \emph{Performance Evaluation}, vol 66: 3, pp. 621 -- 639.

\bibitem{Boxma11}
Boxma, O.J., Kella, O., \& Kosi\'{n}ski. (2011). \newblock ``Queue Lengths and Workloads in Polling Systems'' \newblock \emph{Operations Research Letters}, vol 39: 6, pp. 401 – 405.

\bibitem{RBCooper96}
Cooper, R.B., Niu, S.C., \& Srinivasan, M.M. (1996). \newblock ``A Decomposition Theorem for Polling Models: The Switchover Times are Effectively Additive'', \newblock \emph{Operations Research}, vol 44: 4, pp. 629 -- 633.

\bibitem{Durrett11}
Durrett, R. (2011). \newblock ``Essentials of Stochastic Processes'', \emph{Springer-Verlag New York}, pp. 133 -- 136.

\bibitem{Frigui97}
Frigui, I. (1997). \newblock ``Analysis of Time-Limited Polling System with Markovian Arrival Process and Phase Type Service'', \newblock \emph{PhD Thesis}. University of Manitoba, Manitoba, Canada.

\bibitem{Fuhrmann85}
Fuhrmann, S.W., \& Cooper, R.B. (1985). \newblock ``Stochastic Decomposition in the M/G/1 Queue with Generalized Vacations'', \newblock \emph{Operations Research}, vol 33: 5, pp. 1117 -- 1129.

\bibitem{Lawler06}
Lawler, G.F. (2006). \newblock ``Introduction to Stochastic Process'', \emph{Taylor \& Francis Group}, pp. 26 -- 30.

\bibitem{Prabhu60}
Prabhu, N.U. (1960). \newblock ``Some Results for the Queue with Poisson Arrivals'', \newblock \emph{Journal of the Royal Statistical Society. Series B (Methodological)}, vol 22: 1, pp. 104 -- 107.

\bibitem{Resing93}
Resing, J.A.C.  (1993). \newblock ``Polling systems and multitype branching processes'', \newblock \emph{Queueing Systems}, vol 13: 3, pp. 409 -- 426.

\bibitem{Srinivasan95}
Srinivasan, M.M., Niu, S.C., \& Cooper, R.B. (1995). \newblock ``Relating Polling Models With Zero and Nonzero Switchover Times'', \newblock \emph{Queueing Systems}, vol 19: 1, pp. 149 -- 168.

\bibitem{Takagi90}
Takagi, H., (1990). \newblock ``Queueing Analysis of Polling Models: An Update'', \emph{Stochastic Analysis of
Computer and Communication Systems}, North-Holland, Amsterdam, pp. 267 -- 318.

\bibitem{Takagi2000}
Takagi, H. (2000). \newblock ``Analysis and Application of Polling Models. In: Haring G., Lindemann C., Reiser M. (eds)'', \newblock \emph{Performance Evaluation: Origins and Directions. Lecture Notes in Computer Science}, vol 1769, pp. 423 -- 442, Springer, Berlin, Heidelberg.

\bibitem{Vishnevskii2006}
Vishnevskii, V.M., \& Semenova, O.V. (2006). \newblock ``Mathematical Methods to Study the Polling Systems'', \newblock
\emph{Automation and Remote Control}, vol 67: 2, pp. 173 -- 220.

\bibitem{Winands06}
Winands, E.M.M., Adan, I.J.B.F. \& van Houtum, G. (2006). \newblock ``Mean Value Analysis for Polling Systems'' \newblock \emph{Queueing Systems}, vol 54: 35, pp. 35 – 44.

\bibitem{Winands11}
Winands, E.M.M. (2011). \newblock ``Branching-Type Polling Systems with Large Setups'' \newblock \emph{OR Spectrum}, vol 33: 1, pp. 77 – 97.
\end{thebibliography}
\end{document}